\documentclass{JHEP3}
\usepackage{epsf,amsmath,amssymb,scalefnt,rotating,cite}
\usepackage{graphicx}
\newcommand{\cew}[1]{c_{#1}^{\rm EW}}
\newcommand{\bare}{{\rm B}}
\newcommand{\ds}{\displaystyle}

\newcommand{\abbrev}{\scalefont{.9}}
\newcommand{\ep}{\epsilon}
\newcommand{\api}{\frac{\alpha_s}{\pi}}
\newcommand{\eqn}[1]{Eq.\,(\ref{#1})}
\newcommand{\fig}[1]{Fig.\,\ref{#1}}

\newcommand{\sct}[1]{Sect.\,\ref{#1}}
\newcommand{\appx}[1]{App.\,\ref{#1}}
\newcommand{\dd}{{\rm d}}

\newcommand{\order}[1]{{\cal O}(#1)}
\newcommand{\lo}{{\abbrev LO}}
\newcommand{\nlo}{{\abbrev NLO}}
\newcommand{\nnlo}{{\abbrev NNLO}}
\newcommand{\msbar}{{\overline{\rm\abbrev MS}}}
\newcommand{\drbar}{{\overline{\rm\abbrev DR}}}
\newcommand{\dred}{{\abbrev DRED}}
\newcommand{\dreg}{{\abbrev DREG}}
\newcommand{\bld}[1]{\boldmath{$#1$}}
\newcommand{\subhk}{}

\newcommand{\sm}{Standard Model}
\newcommand{\qcd}{{\abbrev QCD}}
\newcommand{\mssm}{{\abbrev MSSM}}
\newcommand{\susy}{{\abbrev SUSY}}
\newcommand{\gfermi}{G_{\rm F}}
\newcommand{\higgs}{h}
\newcommand{\sprod}[2]{#1\!\cdot\!#2}
\newcommand{\Ltop}{L_t}
\newcommand{\Lgluino}{L_{\tilde g}}
\newcommand{\Lstop}[1]{L_{\tilde t#1}}
\newcommand{\mtop}{m_t}
\newcommand{\mstop}[1]{m_{\tilde t_{#1}}}

\newcommand{\mgluino}{m_{\tilde g}}
\newcommand{\muSUSY}{\mu_{\rm \susy{}}}
\newcommand{\muR}{\mu_{\rm R}}

\newcommand{\gghsoftvirt}{Harlander:2000mg,Harlander:2001is,Catani:2001ic}
\newcommand{\gghnnlo}{Harlander:2002wh,Anastasiou:2002yz,Ravindran:2003um}
\newcommand{\nlotth}{Beenakker:2001rj,Dawson:2002tg,Dawson:2003zu}
\newcommand{\nlowbf}{Han:1991ia,Figy:2003nv}
\newcommand{\gghnlo}{Graudenz:1992pv,Spira:1995rr}
\newcommand{\gghnloeff}{Dawson:1990zj,Djouadi:1991tk}

\def\readRCS$#1: #2,v #3 #4 #5${%
 \def\filename{#2}%
 \def\fileversion{#3}%
 \def\filedate{#4}%
}

\readRCS
 $Id: gghsusy.tex,v 1.25 2004/09/20 09:05:16 rharland Exp $

\title{  Supersymmetric Higgs production in gluon fusion at
  next-to-leading order }
\author{Robert V. Harlander\\
  {\it Institut f\"ur Theoretische Teilchenphysik,
  Universit\"at Karlsruhe\\
  D-76128 Karlsruhe, Germany}\\
    E-mail: \email{robert.harlander@cern.ch}}
\author{Matthias Steinhauser\\
  {\it II. Institut f\"ur Theoretische Physik,
  Universit\"at Hamburg\\
  D-22761 Hamburg, Germany}\\
  E-mail: \email{matthias.steinhauser@desy.de}}

\preprint{\hepph{0409010}, DESY 04--150, TTP 04--19, SFB/CPP-04-37 ---
  September 2004}
\date{}

\abstract{
The next-to-leading order (\nlo{}) \qcd{} corrections to the production
and decay rate of a Higgs boson are computed within the framework of the
Minimal Supersymmetric Standard Model (\mssm{}). The calculation is
based on an effective theory for light and intermediate mass Higgs
bosons.  We provide a {\tt Fortran} routine for the numerical evaluation
of the coefficient function.  For most of the \mssm{} parameter space,
the relative size of the \nlo{} corrections is typically of the order of
5\% smaller than the \sm{} value. We exemplify the numerical results for
two scenarios: the benchmark point {\abbrev SPS}\,1a, and a parameter
region where the gluon-Higgs coupling at leading order is very small due
to a cancellation of the squark and quark contributions.
}

\begin{document}

\section{Introduction}

The most important Higgs production cross sections at the Tevatron and
the Large Hadron Collider ({\abbrev LHC}) are known quite reliably
within the Standard Model framework. Weak boson fusion as well as
associated production of Higgs bosons with a top quark pair are known
with \nlo{} \qcd{} accuracy (see Refs.~\cite{\nlowbf} and
\cite{\nlotth}, respectively), while gluon fusion and Higgs-Strahlung
are even known through \nnlo{} in \qcd{} (see
Refs.~\cite{\gghnloeff,\gghnlo,\gghsoftvirt,\gghnnlo} and
Ref.~\cite{Brein:2003wg}). At this level of accuracy, also electro-weak
corrections can be important. They have been evaluated to first order
for the Higgs-Strahlung process~\cite{Ciccolini:2003jy} (a combined
analysis of \qcd{} and electro-weak effects can be found in
\cite{Brein:2004ue}). For the gluon fusion process, only very recently
the full set of first order electro-weak effects has been completed (for
$M_h<2M_W$): The contribution from light fermion loops at two-loop order
was evaluated in Ref.~\cite{Aglietti:2004nj}, the top quark induced
effects were calculated in Ref.~\cite{Degrassi:2004mx}.  The terms of
order $\gfermi \mtop^2$ have been known for ten
years~\cite{Djouadi:1994ge}, and additional \qcd{} effects of order
$\alpha_s \gfermi \mtop^2$ can be extracted from the result of
Ref.~\cite{Steinhauser:1998cm}.

One remarkable fact about the gluon fusion process is that it has no
tree-level contribution. It is therefore sensitive to new particles that
can mediate the gluon-Higgs coupling. In the Standard Model, this
coupling is dominated by a top quark loop, with a small contribution
also from bottom quarks. In the \mssm{}, the bottom quark contribution
can be enhanced for large values of $\tan\beta$. Their \nlo{} effects
have been evaluated in Refs.~\cite{\gghnlo}.

In this paper, we consider the gluon-Higgs coupling mediated by quarks
and squarks at \nlo{} and its effects on the hadronic production and
decay of a light or intermediated mass scalar Higgs boson.  A
preliminary study of these effects has been published in
Ref.~\cite{Harlander:2003bb}, where, however, only a very restricted
parameter range of the \mssm{} has been used. In particular, mixing in
the stop sector was neglected. In this work we dismiss these
constraints. In combination with Refs.~\cite{\gghnlo} for the bottom
loops, we thus provide the \nlo{} result for almost all of the \susy{}
parameter space. The only restrictions are in the region where
sbottom-effects become important. However, this only happens when the
sbottom mixing angle is $\theta_b\approx 45^\circ$ and both the mass
splitting between $\tilde b_1$ and $\tilde b_2$ as well as $\tan\beta$
are very large.

The outline of the paper is as follows. In \sct{sec::effnlo}, we
construct the effective Lagrangian for the gluon-Higgs interaction by
integrating out the top quark as well as all supersymmetric particles of
the \mssm{}. The Lagrangian at leading order in $1/M$ ($M\in
\{\mtop,\mstop{},\mgluino\}$) contains only one operator. We evaluate
its universal Wilson coefficient through \nlo{}.  In \sct{sec::results},
we first study the behavior of the \nlo{} Wilson coefficient itself, and
subsequently use it to evaluate the hadronic decay and production rate
of the light, {\abbrev CP}-even Higgs boson of the \mssm{}. To this aim,
we choose two specific sets of \susy{} parameters: The first one is
the benchmark {\abbrev SPS}\,1a, defined in Ref.~\cite{Allanach:2002nj}. The
second one is similar to the ``gluophobic Higgs'' scenario
of Ref.~\cite{Carena:1999xa}, where the squark and the quark contributions to
the gluon-Higgs coupling nearly cancel each
other~\cite{Djouadi:1998az}. In \sct{sec::conclusions}, we present our
conclusions.  \appx{sec::frules} and \ref{sec::renorm} collect the
Feynman rules, counter terms, and decoupling constants that we used in
our calculation. \appx{sec::evalcsusy} describes the {\tt Fortran}
routine that we provide to evaluate the first and second order Wilson
coefficient.

\section{Effective Lagrangian through next-to-leading order}\label{sec::effnlo}

\subsection{Effective Lagrangian}

The effective Lagrangian is constructed from the full \mssm{} Lagrangian
by integrating out all \susy{} partners and the top quark. The field
content of the effective theory is thus the same as when starting from
the \sm{} Lagrangian. Therefore, also the effective Lagrangian has the
same form. It is given by
\begin{equation}
\begin{split}
{\cal L}_{\rm eff} &= -\frac{\higgs}{v}\,C_1^\bare\,{\cal
  O}_1^\bare\,,\qquad
{\cal O}_1^\bare = \frac{1}{4}G_{a,\mu\nu}^\bare G_a^{\bare,\mu\nu}\,,
\label{eq::efflag}
\end{split}
\end{equation}
where $G_{a,\mu\nu}^\bare$ is the bare gluonic field strength tensor,
$v\approx 246$\,GeV, and $C_1^\bare$ is the matching coefficient to the
full theory.  For the sake of simplicity of the discussion, we focus on
the light neutral Higgs, denoted $\higgs$, in this paper. The
translation of the formulas to the heavy neutral Higgs is
straightforward, but the validity of the effective theory approach of
\eqn{eq::efflag} has to be carefully checked in this case.  

For the \sm{}, the two-loop $\alpha_s^2$ corrections for $C_1(\alpha_s)$
have been calculated in Ref.~\cite{\gghnloeff}, the $\alpha_s^3$ and
$\alpha_s^4$ terms in Ref.~\cite{Chetyrkin:1997iv,Kramer:1996iq} and
Ref.~\cite{Chetyrkin:1997un}, respectively.  Furthermore, as mentioned
in the introduction, electroweak corrections of order $\gfermi \mtop^2$
and $\alpha_s \gfermi \mtop^2$ have been evaluated
Refs.~\cite{Djouadi:1994ge} and~\cite{Steinhauser:1998cm}, respectively.
For the \mssm{}, the two-loop \qcd{} corrections are known in the case
of zero squark mixing~\cite{Harlander:2003bb}.

The \qcd{} renormalization of $C_1^\bare$ and ${\cal O}_1^\bare$ is
discussed in Refs.~\cite{Spiridonov:1984br,Chetyrkin:1997un} and is
given by
\begin{equation}
\begin{split}
C_1 &= Z_{11}^{-1}\,C_1^\bare\,,\quad
{\cal O}_1 = Z_{11}\,{\cal O}_1^\bare\,,\quad\mbox{with}\quad
Z_{11} = \left(1 - \frac{\pi}{\alpha_s}\frac{\beta}{\ep}\right)^{-1}\,,
\label{eq::opren}
\end{split}
\end{equation}
where 
\begin{equation}
\begin{split}
\beta(\alpha_s) = -\left(\frac{\alpha_s}{\pi}\right)^2\,\beta_0 +
\order{\alpha_s^3}
\end{split}
\end{equation}
is the $\beta$-function of standard $(n_l=5)$-flavor \qcd{}, with
$\beta_0 = 11/4 - n_l/6$. Note that here and in what follows, $\alpha_s$
denotes the strong coupling constant in standard five-flavor \qcd{}; it
is a function of the renormalization scale $\muR$:
\begin{equation}
\begin{split}
\alpha_s \equiv \alpha_s^{(5)}(\muR)\,.
\end{split}
\end{equation}

In this paper we calculate $C_1$ in the \mssm{} through $\alpha_s^2$,
i.e., we will evaluate the coefficients $c_1^{(0)}$ and $c_1^{(1)}$
defined as
\begin{equation}
\begin{split}
C_1 &= -\frac{\alpha_s}{3\pi}\left[ c_1^{(0)} + \api c_1^{(1)} +
  \order{\alpha_s^2} \right]\,.
\end{split}
\label{eq::C1}
\end{equation}
This will allow us to compute the \nlo{} approximation to the hadronic
production and decay rate of a {\abbrev CP}-even Higgs boson.  Following
the argumentation of Ref.~\cite{Harlander:2003kf}, we can even derive a
fairly accurate estimate of the \nnlo{} production cross section in this
model.

Several methods to compute the coefficient function $C_1$ are described
in Ref.~\cite{Steinhauser:2002rq}. Here we follow the most direct one which is
based on the relation
\begin{equation}
\begin{split}
\zeta_3^\bare C_1^\bare
&=-\frac{1}{4}\left(
\frac{g^{\mu\nu}(\sprod{p_1}{p_2}) - p^\nu_{1}\,p^\mu_{2} 
  - p^\mu_{1}p^\nu_{2}}%
{(D-2)(\sprod{p_1}{p_2})^2}\,
\Gamma^{\bare}_{\mu\nu}(p_1,p_2)\right)\bigg|_{p_1=p_2=0}\,.
\label{eq::projector}
\end{split}
\end{equation}
$\Gamma^{\bare}_{\mu\nu}(p_1,p_2)$ is the 1-particle-irreducible vertex
function of two gluons in a color-singlet state (incoming momenta $p_1$,
$p_2$) and a Higgs boson in the full theory.  $\zeta_3^\bare$ is the
decoupling constant that relates the gluon field in the full and the
effective theory (details can be found in
Ref.~\cite{Steinhauser:2002rq}). It can be computed from the gluon
propagator in the full theory $\Pi_g^\bare(p)$ through
\begin{equation}
\begin{split}
\zeta_3^\bare &= 1 + \Pi_g^\bare(p=0)\,.
\end{split}
\end{equation}
The result for $\zeta_3^\bare$ is given in \eqn{eq::zeta30}.

\FIGURE{
    \begin{tabular}{ccc}
      \includegraphics[bb=154 450 450
      660,width=8em]{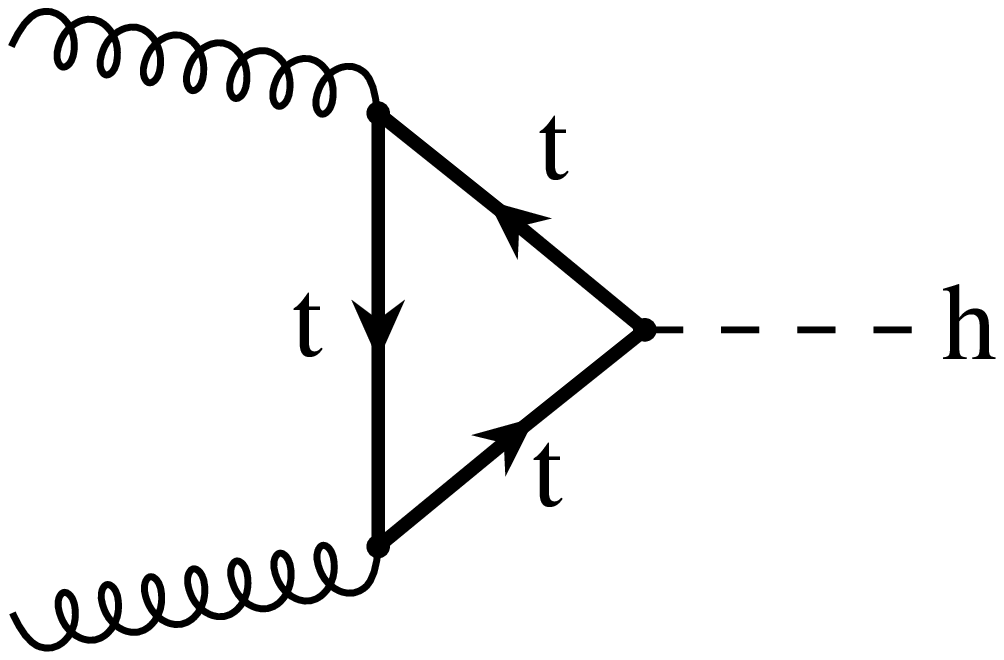} &
      \includegraphics[bb=154 450 450
      660,width=8em]{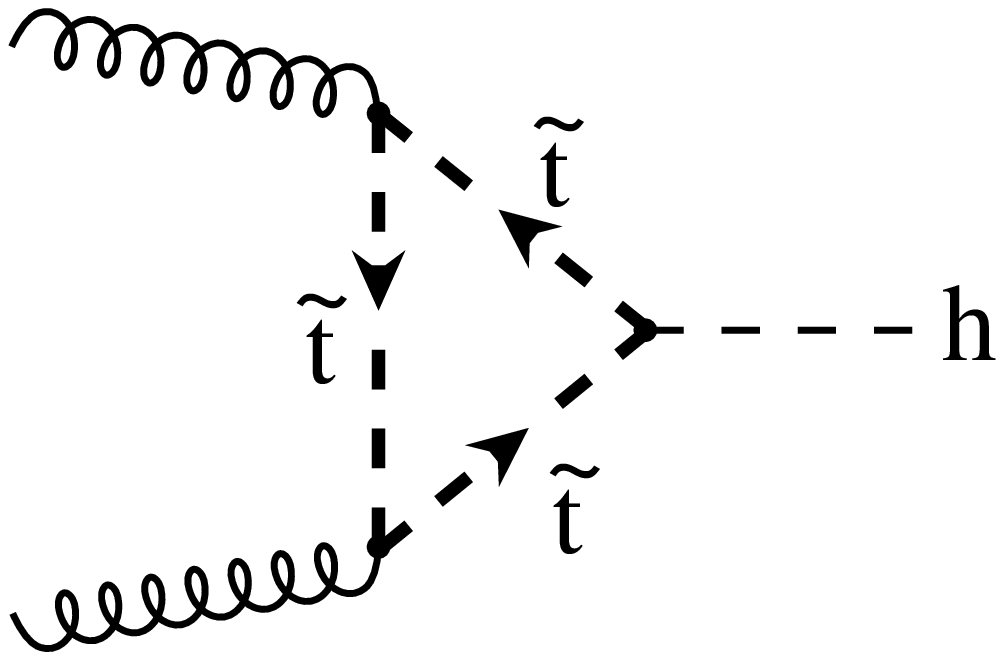} &
      \includegraphics[bb=154 450 450
      660,width=8em]{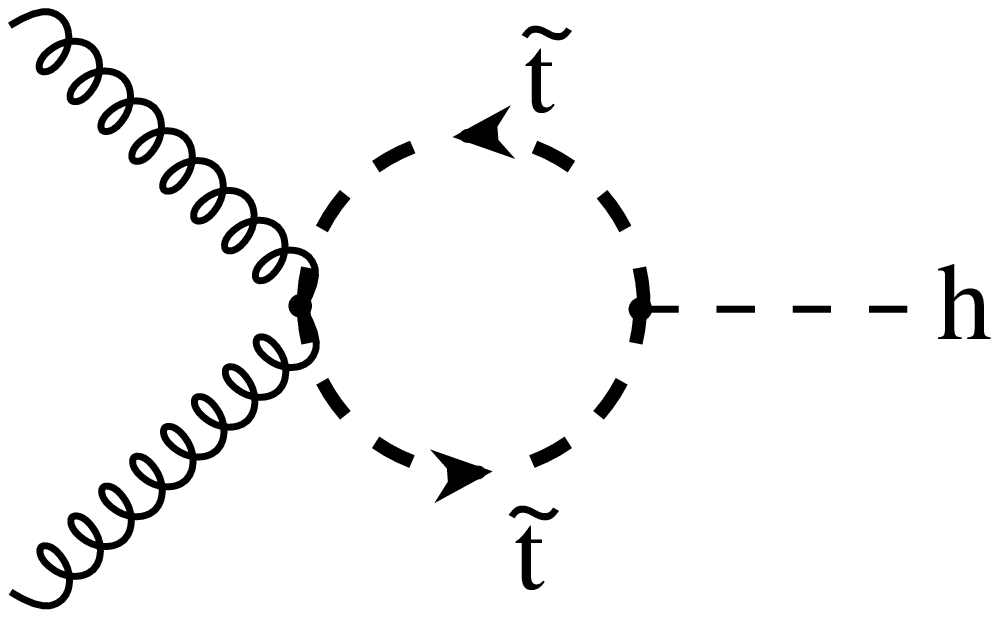} \\
      (a) & (b) & (c) \\[2em]
      \includegraphics[bb=154 450 450
      660,width=8em]{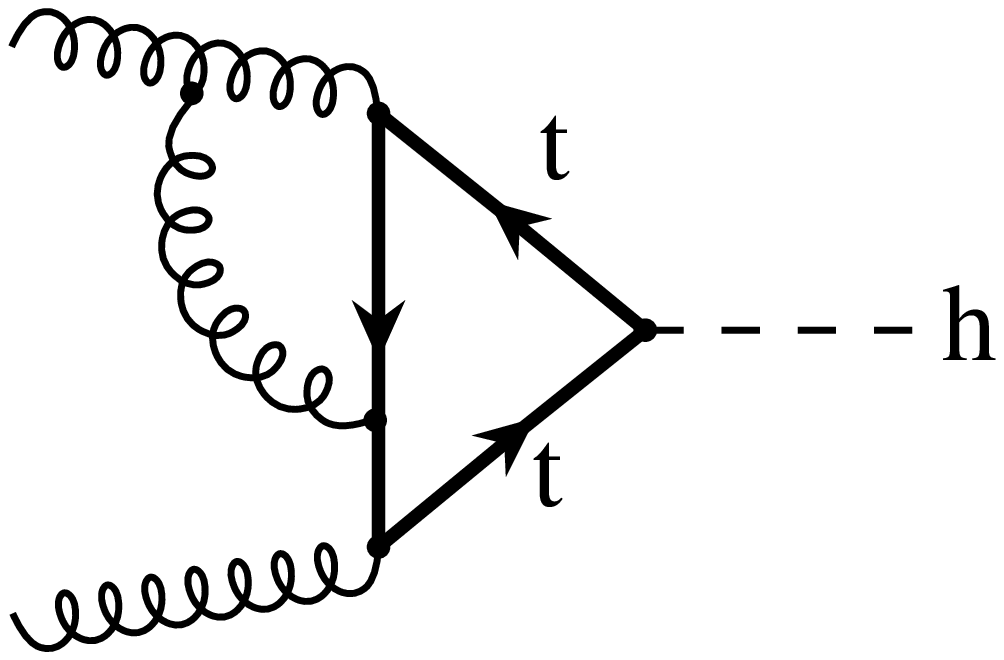} &
      \includegraphics[bb=154 450 450
      660,width=8em]{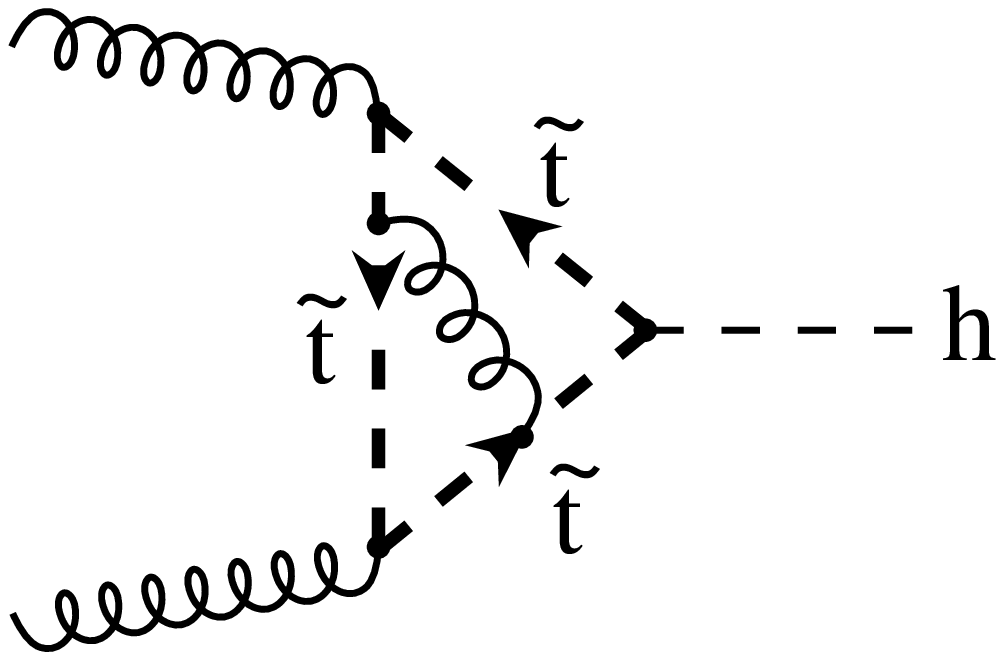} &
      \includegraphics[bb=154 450 450
      660,width=8em]{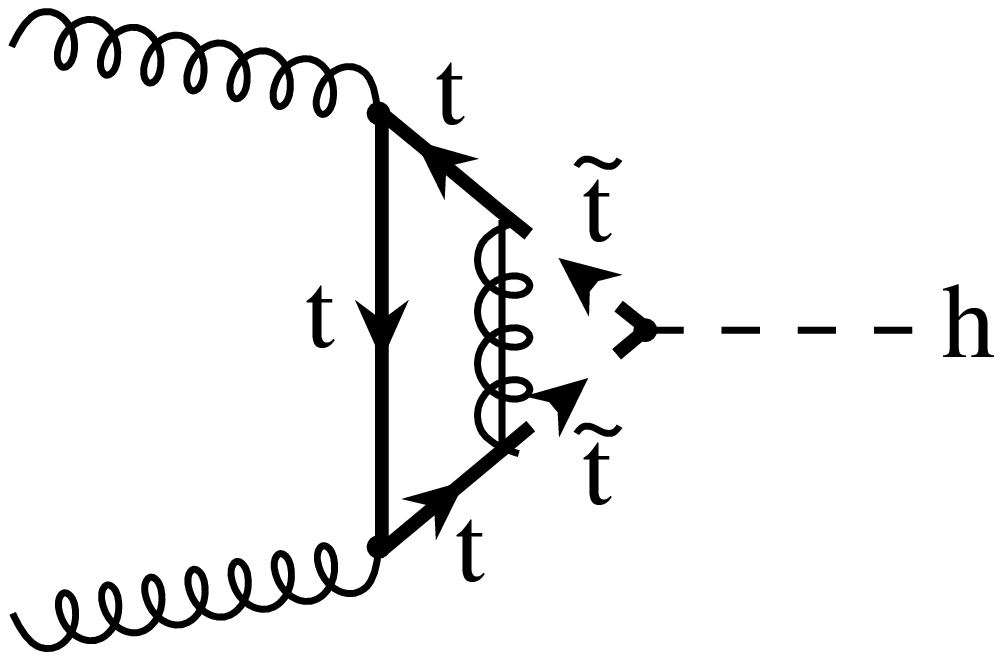} \\
      (d) & (e) & (f) \\[2em]
      \includegraphics[bb=154 450 450
      660,width=8em]{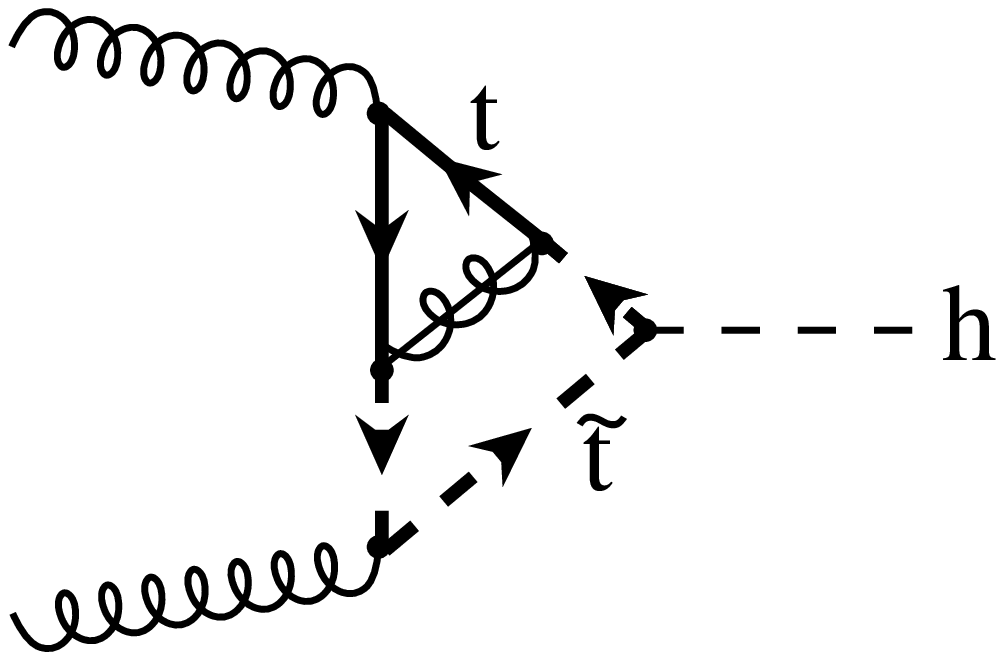} &
      \includegraphics[bb=154 450 450
      660,width=8em]{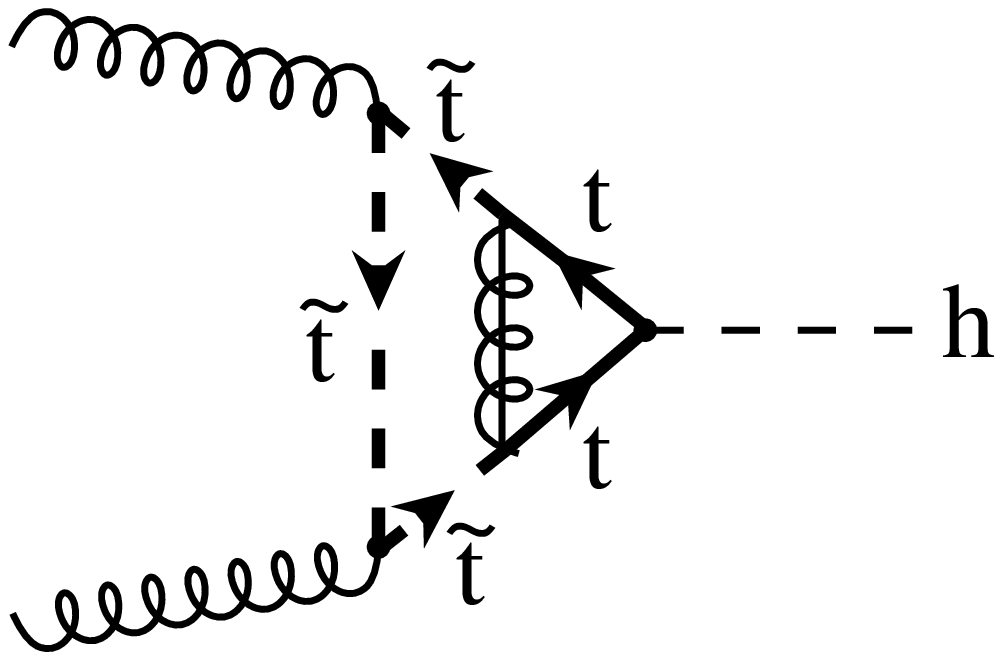} &
      \includegraphics[bb=154 450 450
      660,width=8em]{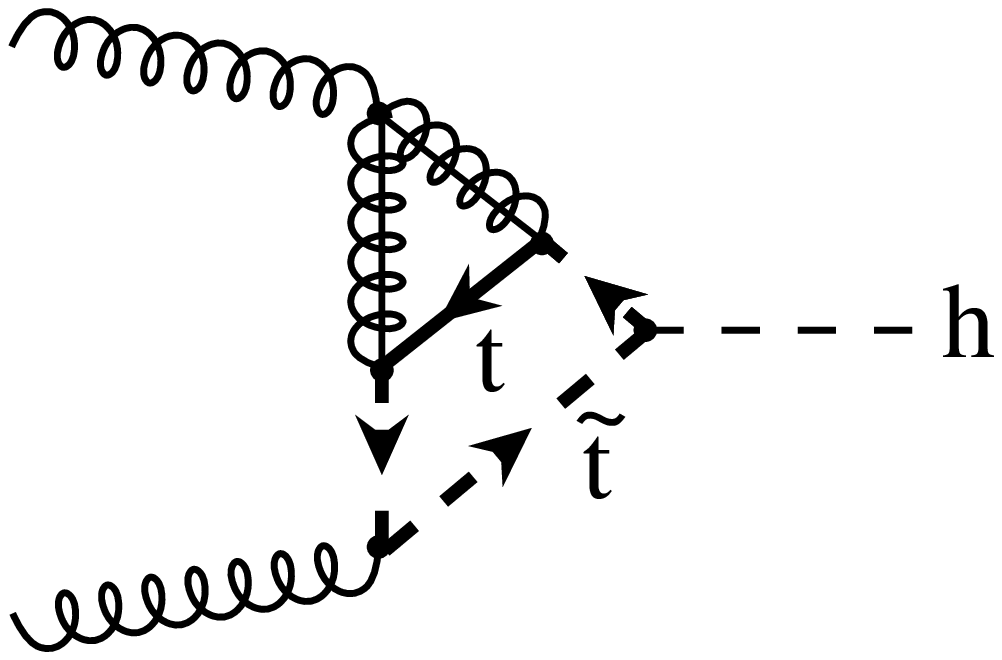}\\
      (g) & (h) & (i)
    \end{tabular}
    \parbox{14.cm}{
      \caption[]{\sloppy
Diagrams contributing to the effective $gg\higgs$ coupling in the
\mssm{}.\label{fig::dias}
        }
}
}

Sample diagrams corresponding to $\Gamma^{\bare}_{\mu\nu}$ are shown in
\fig{fig::dias}.  We may distinguish three different types:
\begin{enumerate}
\item pure top contributions, e.g.\ \fig{fig::dias}\,(a) and (d)
\item pure stop contributions, e.g.\ \fig{fig::dias}\,(b), (c), and (e)
\item mixed top/stop/gluino contributions, e.g.\ \fig{fig::dias}\,(f)-(i)
\end{enumerate}
The pure top quark contributions are separately finite (they correspond 
to the \sm{} terms), while the pure stop and the mixed contributions each
develop ultra-violet poles that cancel in their sum (after 
taking into account the proper counter terms).
Application of \eqn{eq::projector} leads to one- and two-loop integrals with
vanishing external momenta. They can be evaluated in closed form using
the algorithm of Davydychev and Tausk~\cite{Davydychev:1992mt}. Details
will be given in \sct{sec::nlo}.

\subsection{Leading order coefficient function}
The \lo{} approximation of the coefficient function is obtained from the
one-loop diagrams of \fig{fig::dias}\,(a), (b), and (c). The
result is
\begin{equation}
\begin{split}
c_1^{(0)} &= c^{(0)}_{1,t} + c^{(0)}_{1,\tilde t}\,,\qquad
c^{(0)}_{1,t} = \frac{\cos\alpha}{\sin\beta}\,,\\
c^{(0)}_{1,\tilde t} &=
 \frac{\cos\alpha}{\sin\beta}
\bigg[
\frac{1}{4}\left(\frac{\mtop^2}{\mstop{1}^2} 
                           + \frac{\mtop^2}{\mstop{2}^2}\right)
  + \frac{\sin^2{2\theta_t}}{8}\left(1 - \frac{\mstop{1}^2}{2\mstop{2}^2}
 - \frac{\mstop{2}^2}{2\mstop{1}^2} \right)
\bigg]\\
& +
\frac{1}{8}\muSUSY{}\,\mtop\,\frac{\cos(\alpha-\beta)}{\sin^2\beta}\,
\sin{2\theta_t}\,\bigg[
\frac{1}{\mstop{1}^2} - \frac{1}{\mstop{2}^2}
\bigg] \\
&+
\frac{\cew{1}+\cew{2}}{16}\bigg(
  \frac{\mtop^2}{\mstop{1}^2}+\frac{\mtop^2}{\mstop{2}^2}\bigg)
 + \frac{\cew{1}-\cew{2}}{16}\,\cos{2\theta_t}\,\bigg(
  \frac{\mtop^2}{\mstop{1}^2}-\frac{\mtop^2}{\mstop{2}^2}\bigg)\,.
\label{eq::c1lo}
\end{split}
\end{equation}
Here and in the following we assume all masses in the on-shell scheme,
if not stated otherwise ($\mtop$, $\mstop{i}$, and $\mgluino$ are the
top, stop ($i=1,2$) and gluino mass, respectively).  $\alpha$ is the
mixing angle between the weak and the mass eigenstates of the neutral
scalar Higgs bosons, $\tan\beta$ is the ratio of the vacuum expectation
values of the two Higgs doublets, $\muSUSY$ is the coefficient of the
bilinear Higgs term in the \mssm{} superpotential, $\cew{1}$ and
$\cew{2}$ are defined in \eqn{eq::cew}, and the mixing angle $\theta_t$
between the helicity ($\tilde t_L,\tilde t_R$) and mass eigenstates
($\tilde t_1,\tilde t_2$) of the top squarks is defined in
\sct{sec::defs}. For more details on the \mssm{} parameters, see
Ref.~\cite{Martin:1997ns}, for example.  $c^{(0)}_{1,t}$ and
$c^{(0)}_{1,\tilde t}$ are the top- and stop-loop contributions,
respectively. Note that the latter are not necessarily suppressed for
large stop masses. In fact, due to the term $\propto\sin^22\theta_t$,
they can be dominant for large stop mixing and large mass splitting
between $\tilde t_1$ and $\tilde t_2$~\cite{Djouadi:1998az}.  Using
\eqn{eq::thetaqdef}, this term can be written as
\begin{equation}
\begin{split}
\frac{\sin^2 2\theta_t}{8}\left( 1 - \frac{\mstop{1}^2}{2\mstop{2}^2} - 
  \frac{\mstop{2}^2}{2\mstop{1}^2} \right) =
  -X_t^2\frac{\mtop^2}{4\mstop{1}^2\mstop{2}^2}\,,
\end{split}
\label{eq::maxterm}
\end{equation}
implying that the squark effects are large when $X_t = A_t -
\muSUSY\cot\beta$ is large.
In addition, \eqn{eq::c1lo} shows that large effects can also arise from
large values of $\muSUSY$. 

The question arises if cancellations between the quark and squark
contributions occur also when radiative corrections are included, or if
the regions where this occurs are significantly different from the \lo{}
prediction.  We will discuss this issue for a specific example in
\sct{sec::results}.

\subsection{Next-to-leading order coefficient function}\label{sec::nlo}
A major difference between the \sm{} and the \susy{} calculation for
$C_1$ is the occurrence of more than one mass scale in \susy{}; this leads
to expressions for $C_1$ that are much more complicated and unhandy as
compared to the \sm{} result. The latter depends only on the top quark
mass and thus involves only constants and logarithms of the form
$\ln(\muR^2/\mtop^2)$, where $\muR$ is the renormalization scale. In fact, 
let us recall the expression in the
\sm{}~\cite{Chetyrkin:1997iv,Kramer:1996iq}:
\begin{equation}
\begin{split}
  C_1^{\rm SM} &= -\frac{\alpha_s}{3\pi}\left\{1 +
    \frac{11}{4}\api
    + \left[ \frac{2777}{288} + \frac{19}{16}\ln\frac{\muR^2}{\mtop^2} 
      - n_l \left(\frac{67}{96} - \frac{1}{3} \ln\frac{\muR^2}{\mtop^2}\right)
      \right]\left(\api\right)^2
  \right\}
  + \order{\alpha_s^4}\,,
\end{split}
\end{equation}
where for convenience we also displayed the \nnlo{} result.

The calculation of $c^{(1)}_1$ in the \mssm{} leads to two-loop
integrals with up to three different masses $m_1,m_2,m_3\in
\{\mtop,\mstop{1},\mstop{2},\mgluino\}$ (integrals with four different
masses can be transformed to integrals with three different masses by
simple partial fractioning). Davydychev and Tausk have provided an
algorithm for their analytic evaluation~\cite{Davydychev:1992mt}.  It
allows one to express the integrals through the function
\begin{equation}
\begin{split}
\tilde\Phi(m_1,m_2,m_3) &= (m_3\lambda)^2\,\left\{
\begin{array}{ll}
 \Phi_1\left(\frac{m_1^2}{m_3^2},\frac{m_2^2}{m_3^2}
 \right)\,,&\quad m_1+m_2\leq m_3\,,\\[.5em]
 \Phi_2\left(\frac{m_1^2}{m_3^2},\frac{m_2^2}{m_3^2}
 \right)\,,&\quad m_1+m_2 > m_3\,,
\end{array}
\right.
\label{eq::davytausk}
\end{split}
\end{equation}
where
\begin{equation}
\begin{split}
\Phi_1(x,y) &= \frac{1}{\lambda}\,\bigg\{
2\,\ln\left[\frac{1}{2}\left( 1 + x - y - \lambda \right)\right]
\,\ln\left[\frac{1}{2}\left( 1 - x + y - \lambda \right)\right]
- \ln x \ln y\\
&\quad 
- 2\,{\rm Li}_2\left[\frac{1}{2}\left( 1 + x - y - \lambda \right)\right]
- 2\,{\rm Li}_2\left[\frac{1}{2}\left( 1 - x + y - \lambda \right)\right]
+ \frac{1}{3}\pi^2\bigg\}\,,
\end{split}
\end{equation}
and
\begin{equation}
\begin{split}
\Phi_2(x,y) &= 
\frac{2}{\sqrt{-\lambda^2}}\bigg\{
{\rm Cl}_2\left[2\arccos\left(\frac{-1+x+y}{2\sqrt{xy}}\right)\right]\\
&\quad
+{\rm Cl}_2\left[2\arccos\left(\frac{1+x-y}{2\sqrt{x}}\right)\right]
+{\rm Cl}_2\left[2\arccos\left(\frac{1-x+y}{2\sqrt{y}}\right)\right]
\bigg\}\,,
\end{split}
\end{equation}
\begin{equation}
\begin{split}
\lambda &= \sqrt{(1-x-y)^2 - 4xy}\,,\qquad
x = \frac{m_1^2}{m_3^2}\,,\qquad
y = \frac{m_1^2}{m_3^2}\,.
\end{split}
\end{equation}
${\rm Li}_2(x)$ is the standard dilogarithm and ${\rm Cl}_2(x)$ is
Clausen's integral function,
\begin{equation}
\begin{split}
{\rm Li}_2(x) &= -\int_0^1\dd t\,\frac{\ln(1-xt)}{t}\,,\qquad
{\rm Cl}_2(x) = -\int_0^x\dd t\,\ln|2\sin(t/2)|\,.
\end{split}
\end{equation}
Note that $\tilde \Phi(m_1,m_2,m_3)$ is symmetric in $m_1,m_2$ and $m_3$.

To regulate the ultra-violet divergences of the loop integrals, we use
Dimensional Reduction (\dred{}). This is realized by evaluating all
Dirac traces and Lorentz contractions in four, and all loop integrals in
$d=(4-2\ep)$ space-time dimensions~\cite{Jack:1994rk,Jack:1997sr}.  The
external projector defined in \eqn{eq::projector} is taken in $d$
dimensions.  Renormalization is done as explained in \appx{sec::renorm}.

Note that even though we only keep the top- and stop-Higgs couplings
different from zero, squarks of other flavors ($\tilde b$, $\tilde c$
etc.) may enter the \nlo{} calculation through the four-squark vertex
listed in \appx{app::frules}. Typical diagrams are shown in
\fig{fig::hsquarks}. However, diagrams like \fig{fig::hsquarks}\,(a)
vanish due to their color factor, and the diagrams like the one in
\fig{fig::hsquarks}\,(b) add up to zero.
\FIGURE{
    \begin{tabular}{cc}
      \includegraphics[bb=120 520 420 690,width=10em]{%
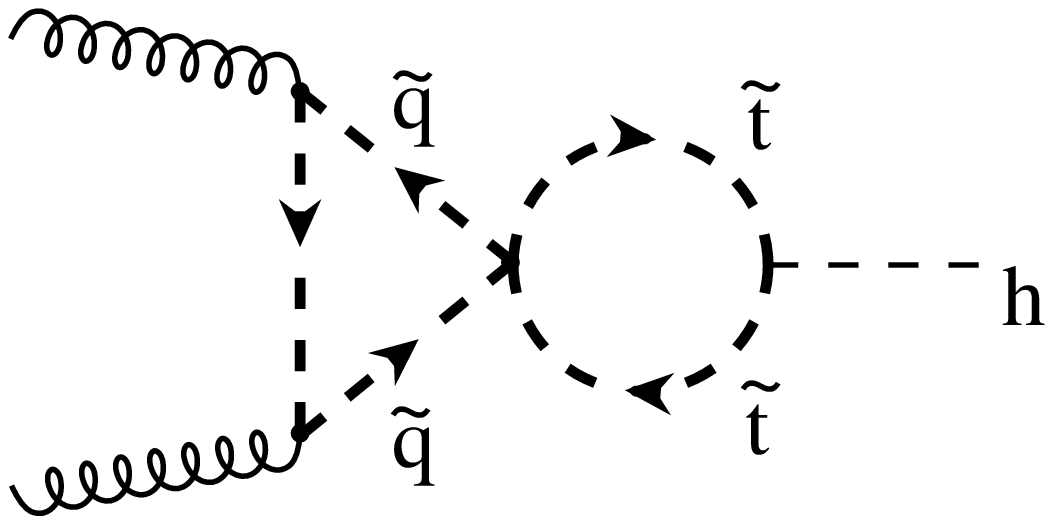}  &\qquad
      \raisebox{-.5em}{\includegraphics[bb=45 520 345 730,width=10em]{%
  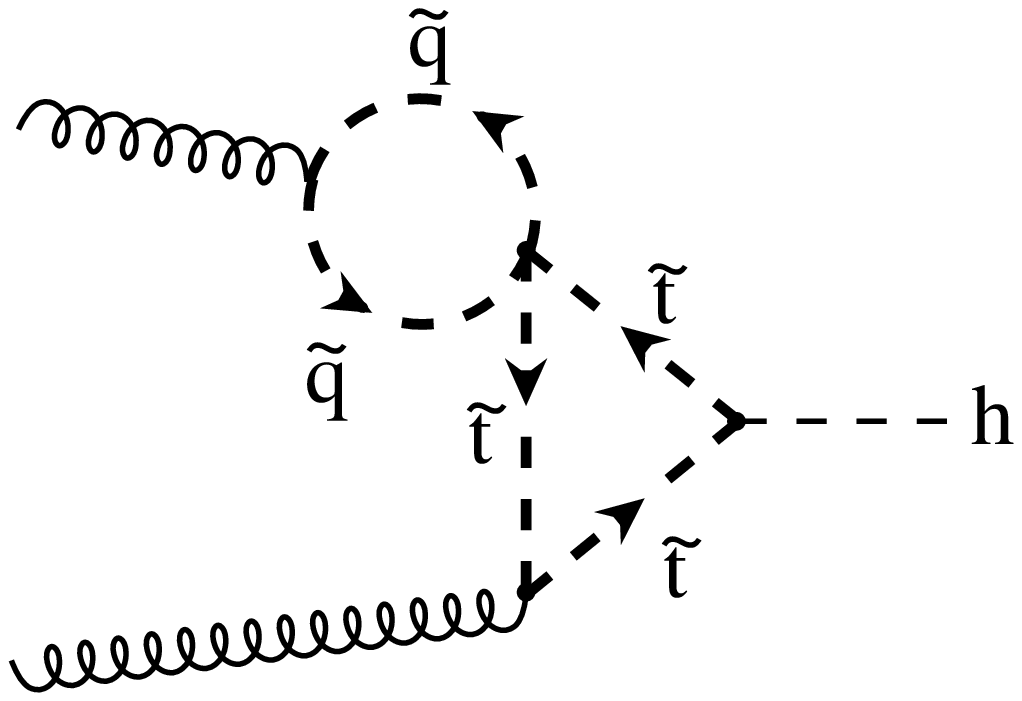} }\\
      (a) & (b)
    \end{tabular}
    \parbox{14.cm}{
      \caption[]{\sloppy
Diagrams involving $\tilde q$, with $q\in\{u,d,c,s,t,b\}$.
\label{fig::hsquarks}
        }}
}

The full result for $c_1^{(1)}$ in the \mssm{} is too long to be quoted
here.\footnote{All algebraic manipulations were done with the help of
{\tt FORM}~\cite{Vermaseren:2000nd}.} Instead, we provide a {\tt
Fortran} code, named {\tt evalcsusy.f}~\cite{evalcsusy}, that allows for
a numerical evaluation of the coefficient function and can be combined
with other programs quite easily using an {\abbrev SLHA}-like interface
({\abbrev SLHA}: \susy{} Les Houches Accord~\cite{Skands:2003cj}).  For
details, see \appx{sec::evalcsusy}.

As a check of our result, we also calculated the diagrams by means of
asymptotic expansions, using the program {\tt EXP}~\cite{exp}.  It
allows us to evaluate an approximate result for $C_1$, provided that
there is a certain hierarchy among the masses
$\mtop,\mstop{1},\mstop{2}$, and $\mgluino$.  The approximation,
however, will only be valid within the radius of convergence of the
specific series, so that we will not make use of it in our
phenomenological analyses below. Needless to say that the expansion of
the analytic result expressed through \eqn{eq::davytausk} agrees with
the corresponding result obtained through asymptotic expansions.

As another check, we reproduced the results of an earlier
publication of ours~\cite{Harlander:2003bb} which was obtained by
asymptotic expansions and in a very simplifying limit.

\section{Results}\label{sec::results}

\subsection[Results for $C_1$]{Results for \bld{C_1}}\label{sec::c1res}

In order to get an impression about the typical size of the corrections
we consider two scenarios. First, we look at the behavior of $c_1^{(0)}$
and $c_1^{(1)}$ at and along a ``Snowmass Point and Slope'' ({\abbrev
SPS})~\cite{Allanach:2002nj}.  In the second case we consider a particular
region of the parameter space where $C_1$ shows large deviations from
its \sm{} value.

To be specific, let us assume an m{\abbrev SUGRA} scenario\footnote{%
Typical {\abbrev GMSB} and {\abbrev AMSB} scenarios as defined by
{\abbrev SPS}\,7 and {\abbrev SPS}\,9~\cite{Allanach:2002nj} give
qualitatively similar results.}  with the five parameters $m_0$,
$m_{1/2}$, $A_0$, $\tan\beta$, and sign$(\muSUSY)$.  In addition, we
define the following \sm{} parameters:
\begin{equation}
\begin{split}
&M_Z = 91.1876\,{\rm GeV}\,,\quad
\bar m_b = 4.2\,{\rm GeV}\,,\quad
\mtop = 178\,{\rm GeV}\,,\quad
m_\tau = 1.777\,{\rm GeV}\,,\\
&\alpha^{-1}_{\rm QED}(M_Z) = 127.934\,,\qquad
\gfermi = 1.16637\cdot 10^{-5}\,{\rm GeV}^{-2}\,,\qquad
\alpha_s(M_Z) = 0.118\,,
\label{eq::sminputs}
\end{split}
\end{equation}
where $M_Z$ is the $Z$ boson mass,
$\bar m_b \equiv \bar m_b(\bar m_b)$ is the scale-invariant
$\msbar{}$ value of the bottom quark mass, $\mtop$ is the
pole mass of the top quark and $m_\tau$ is the mass of the $\tau$ lepton.  
$\alpha_{\rm QED}(M_Z)$ is the running electromagnetic coupling at $M_Z$, 
$\gfermi$ is the Fermi constant and $\alpha_s$ the strong coupling.
As discussed in \appx{sec::renorm}, we further need to
define the (arbitrary) scale $q_0$ (see \eqn{eq::deltatheta}) that
enters the renormalization constant of the stop mixing angle $\theta_t$,
as well as the usual renormalization scale $\muR$. As our default values we
adopt
\begin{equation}
\begin{split}
  q_0 &= \frac{1}{2}\left(\mstop{1} + \mstop{2}\right)\,, \qquad
  \muR = M_h \,.
  \label{eq::q0}
\end{split}
\end{equation}
$M_h$ is the mass of the light {\abbrev CP}-even Higgs boson. Both of
these choices are generally considered to be typical values which
avoid the explicit occurrence of large logarithmic corrections.

As already mentioned above and explained in detail in
App.~\ref{sec::evalcsusy}, the input and output files of the program
{\tt evalcsusy.f} follow the {\abbrev SLHA}
conventions~\cite{Skands:2003cj}.  Among the various \susy{}-spectrum
calculators which are currently
available~\cite{Baer:1999sp,Allanach:2001kg,Djouadi:2002ze,Porod:2003um}
(see, e.g., Ref.~\cite{Allanach:2003jw} for a comparison), only {\tt
SoftSusy}~\cite{Allanach:2001kg} and {\tt SPheno}~\cite{Porod:2003um}
support the {\abbrev SLHA} conventions for both in- and
output. Therefore, we will use these two generators in our analysis.
For our applications they provide almost identical results.

\FIGURE{
    \begin{tabular}{cc}
      \includegraphics[bb=110 265 465 560,width=18em]{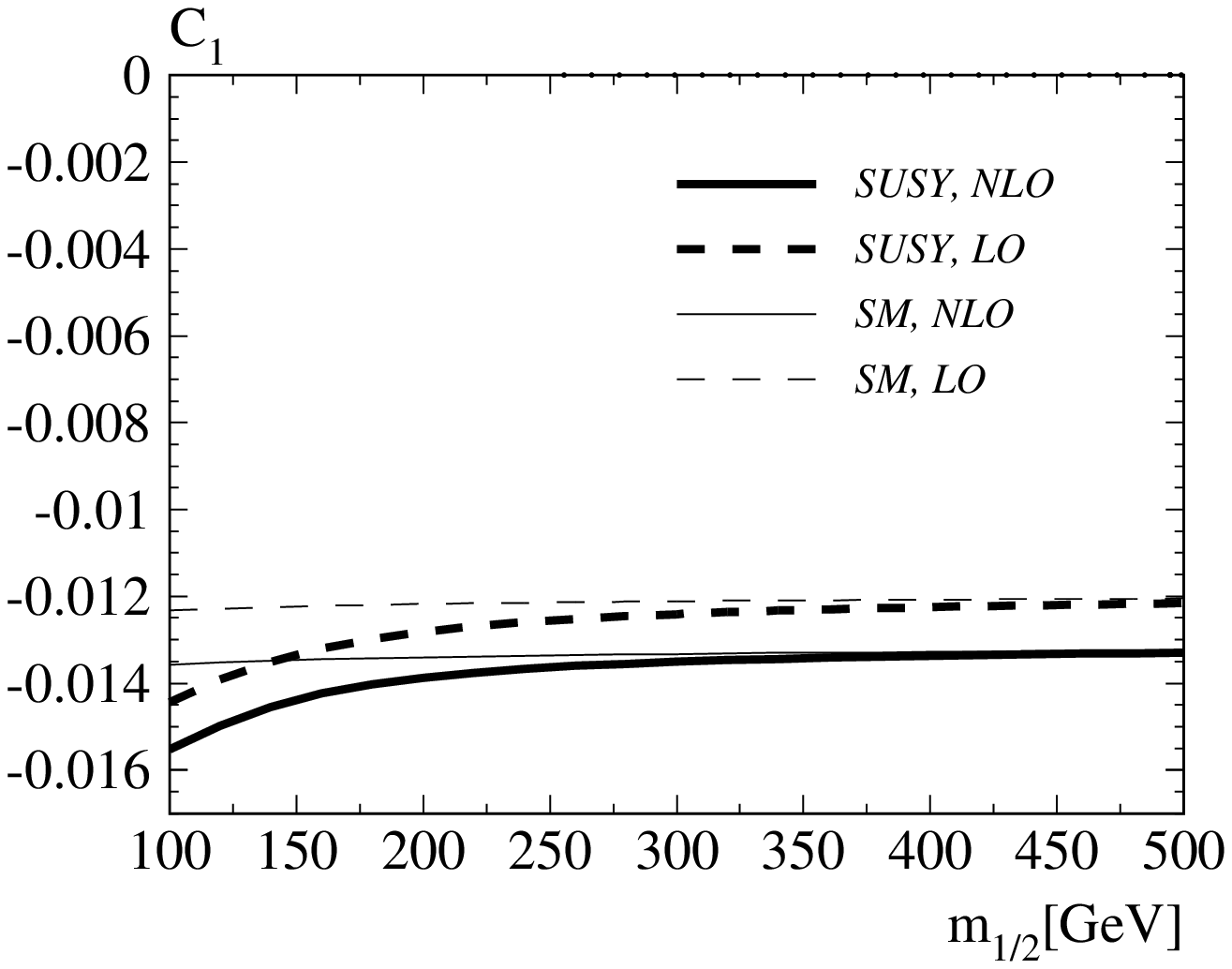}
      \qquad &
      \includegraphics[bb=110 265 465 560,width=18em]{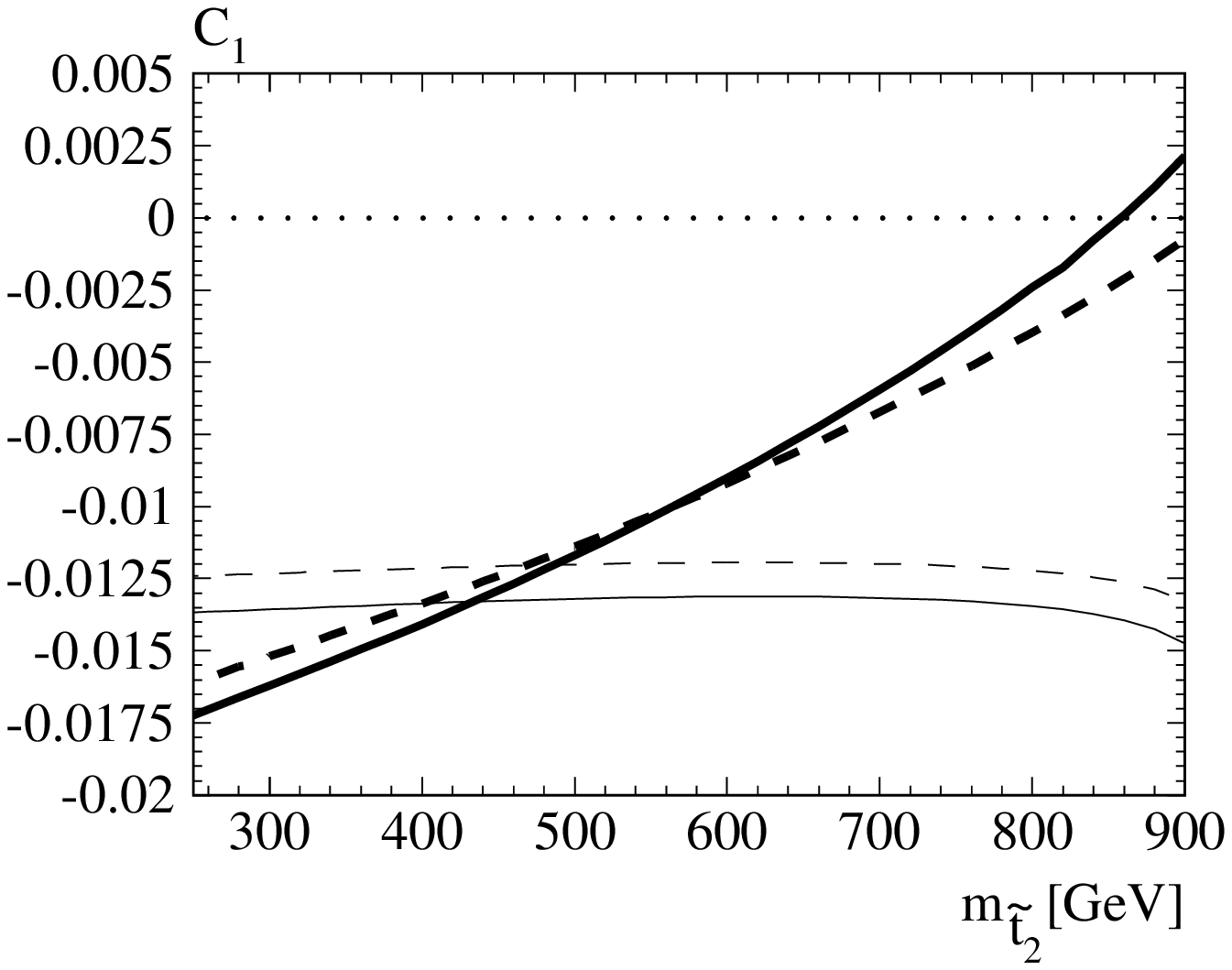} \\
(a) & (b)
    \end{tabular}
    \parbox{14.cm}{
            \caption[]{\label{fig::C1}\sloppy \lo{} (dashed) and \nlo{}
        (solid) result for $C_1$: (a) as a function of $m_{1/2}$ for the
        {\abbrev SPS}\,1a scenario; (b) as a function of $\mstop{2}$ for
        the scenario defined in \eqn{eq::c1max}.  The other parameters
        are fixed as described in the main text.  The corresponding
        Standard Model results are shown as thin lines.}}
}

{\abbrev SPS}\,1a is defined through the following input parameters:
\begin{eqnarray}
  m_0        &=& 100~\mbox{GeV}\,, \nonumber\\
  m_{1/2}    &=& 250~\mbox{GeV}\,, \nonumber\\
  A_0        &=& -100~\mbox{GeV}\,, \nonumber\\
  \tan\beta  &=& 10\,, \nonumber\\
  \mbox{sign}(\muSUSY) &=& +1\,.
  \label{eq::SPS1a}
\end{eqnarray}
Using {\tt SoftSusy} or {\tt SPheno} to derive the low energy parameters
that enter our result and passing them to {\tt evalcsusy.f} as input, 
one finds
\begin{equation}
\begin{split}
  c^{(0)}_1 &\approx 1.03\,,\qquad
  c^{(1)}_1 \approx 2.44\,,
\end{split}
\end{equation}
which is close to the \sm{} values
\begin{equation}
\begin{split}
  c^{(0)}_1 &= 1\,,\qquad
  c^{(1)}_1 = \frac{11}{4} = 2.75\,.
\end{split}
\end{equation}
The slope corresponding to {\abbrev SPS}\,1a is given by
\begin{equation}
\begin{split}
m_0 = -A_0 = 0.4\,m_{1/2}\,,\qquad m_{1/2}\quad \mbox{varies}\,.
\label{eq::sps1aslope}
\end{split}
\end{equation}
The dependence of $C_1$ along this slope is shown as thick lines in
\fig{fig::C1}\,(a) at \lo{} (dashed) and \nlo{} (solid).  One observes a
moderate increase in magnitude of about 8\% when going from \lo{} to
\nlo{}. The thin lines correspond to the \sm{} results. The small
variation of the latter is due to their dependence on $M_h$ through
$\alpha_s(M_\higgs)$. For completeness, let us remark that the masses
that enter our calculation change monotonously within the following
ranges when going from $m_{1/2} = 100$\,GeV to $m_{1/2}=500$\,GeV:
\begin{equation}
\begin{split}
101\,\mbox{GeV} \le\ &M_h \le 118\,\mbox{GeV}\,,\\
176\,\mbox{GeV} \le\ &\mstop{1}\le 784\,\mbox{GeV}\,,\\
330\,\mbox{GeV} \le\ &\mstop{2}\le 1019\,\mbox{GeV}\,,\\
268\,\mbox{GeV} \le\ &\mgluino{}\,\, \le 1158\,\mbox{GeV}\,.
\end{split}
\end{equation}
The dependence of $\mstop{1}$, $\mstop{2}$, and $\mgluino{}$ on
$m_{1/2}$ is almost linear. The dependence of $M_h$ on $m_{1/2}$ is
shown in \fig{fig::mh}\,(a).

\FIGURE{
    \leavevmode
    \begin{tabular}{cc}
      \includegraphics[bb=110 265 465 560,width=18em]{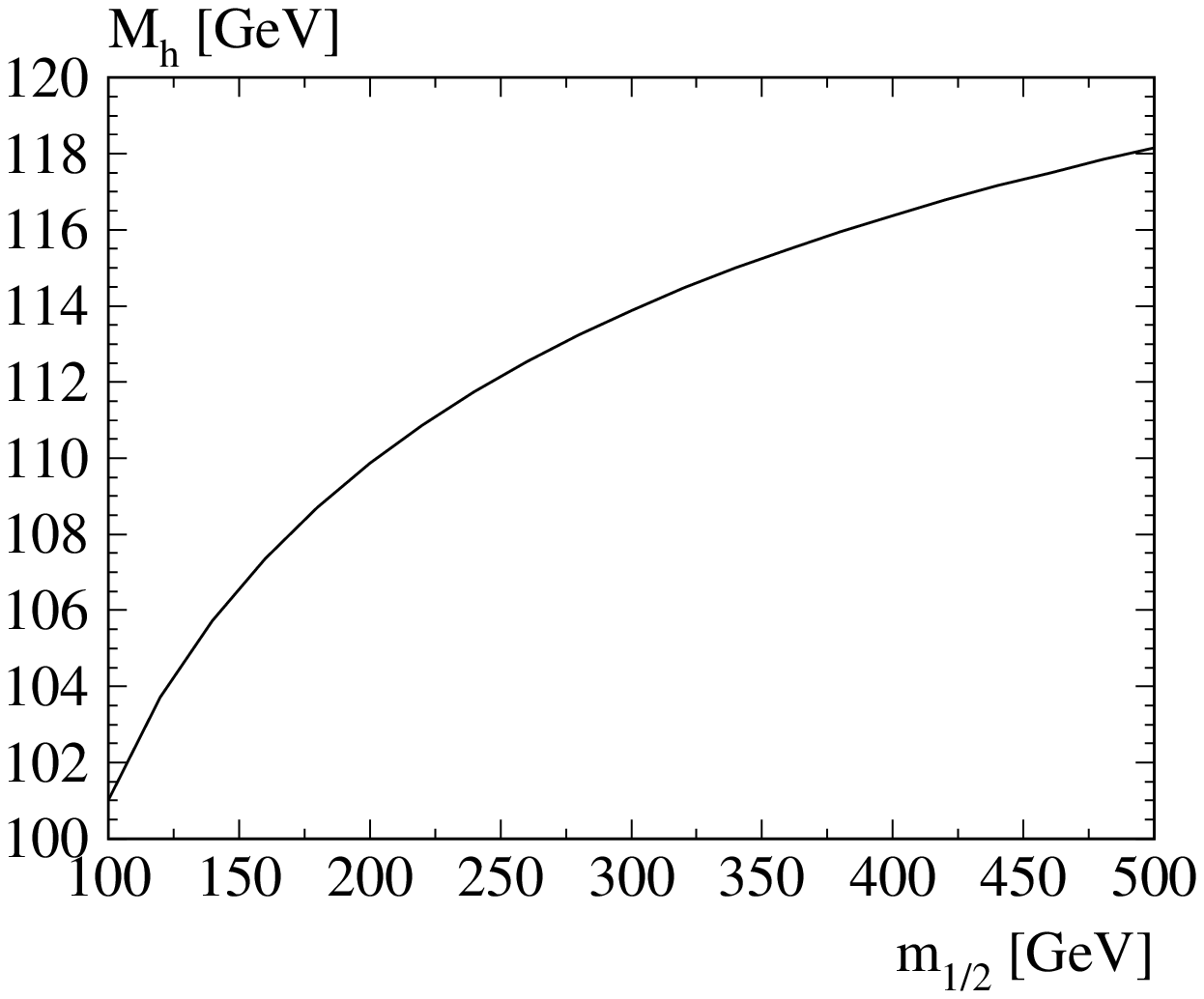} &
      \includegraphics[bb=110 265 465 560,width=18em]{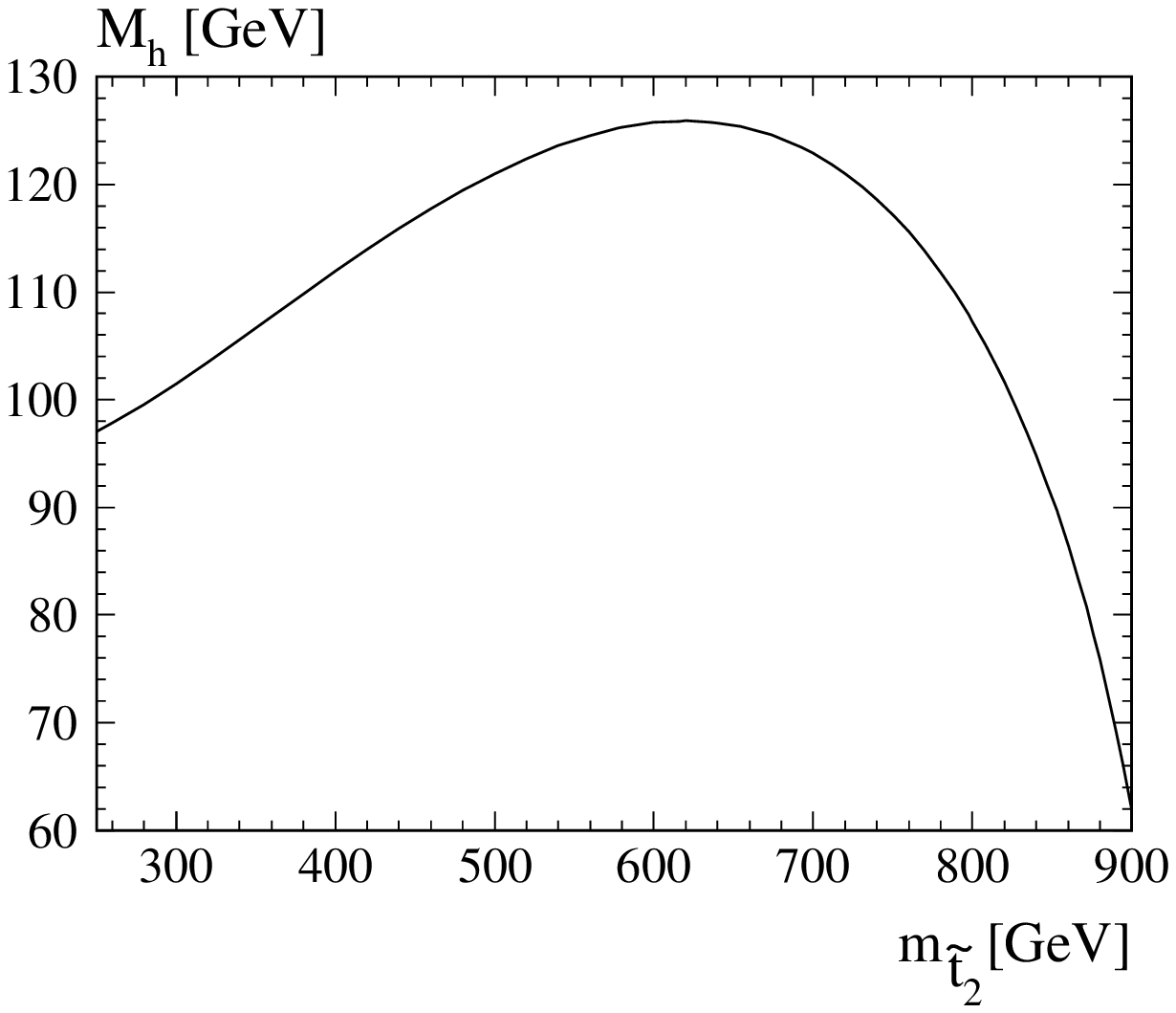} \\
    (a) & (b)
    \end{tabular}
    \parbox{14.cm}{
      \caption[]{\label{fig::mh}\sloppy 
        $M_h$ as a function of (a) $m_{1/2}$ along the slope of 
        {\abbrev SPS}\,1a, and (b) $\mstop{2}$ for the scenario
        defined in \eqn{eq::c1max}.}}
}

In a second example, we consider a case where the \lo{} squark and quark
contributions to the gluon-Higgs coupling largely cancel each
other~\cite{Djouadi:1998az}.  Thus, we do not refer to any \susy{}
breaking scenario, but directly choose the following low energy
parameters:\footnote{ A qualitatively similar benchmark point
(``gluophobic Higgs'') has been suggested in
Ref.~\cite{Carena:2002qg}. }
\begin{equation}
\begin{split}
  &\mtop = 178~\mbox{GeV} \,,\quad \mstop{1} = 200~\mbox{GeV} \,,\quad
  \mgluino = 1000~\mbox{GeV} \,,\quad M_W = 80.425~\mbox{GeV}\,,\\
  &\tan\beta = 10 \,,\qquad \alpha = 0\,,\qquad \theta_t = \frac{\pi}{4}
  \,,\qquad 250\,\mbox{GeV}\le\mstop{2}\le 900\,\mbox{GeV}\,.
  \label{eq::c1max}
\end{split}
\end{equation}
The light Higgs boson mass is determined by the approximate two-loop
formula~\cite{Carena:2000dp}
\begin{equation}
\begin{split}
M_h^2 &= M_Z^2 + M_{h,\alpha}^2 + M_{h,\alpha\alpha_s}^2\,,
\end{split}
\end{equation}
with
\begin{equation}
\begin{split}
M_{h,\alpha}^2 &= \frac{3}{2}\frac{\gfermi\sqrt{2}}{\pi^2}
\left\{-\ln\left(\frac{\mtop^2}{M_S^2}\right)
+ \frac{X_t^2}{M_S^2}\left( 1 - \frac{1}{12}\frac{X_t^2}{M_S^2} \right)
\right\}\,,\\
M_{h,\alpha\alpha_s}^2 &=
-3\,\frac{\gfermi\sqrt{2}}{\pi^2}\frac{\alpha_s}{\pi}
\mtop^4\left\{\ln^2\left(\frac{\mtop^2}{M_S^2}\right)
- \left( 2 + \frac{X_t^2}{M_S^2}
\right)\,\ln\left(\frac{\mtop^2}{M_S^2}\right)
- \frac{X_t}{M_S}\left( 2 - \frac{1}{4}\frac{X_t^3}{M_S^3}\right)
\right\}\,.
\end{split}
\end{equation}
In this approximation and for $\theta_t=\pi/4$, the parameters
$X_t$ and $M_S$ are related to the stop masses through
\begin{equation}
\begin{split}
X_t &= \frac{1}{2\mtop}\left( \mstop{1}^2 - \mstop{2}^2 \right)\,,\qquad
M_S^2 = \frac{1}{2}\left( \mstop{1}^2 + \mstop{2}^2 \right)\,.
\end{split}
\end{equation}
The variation of $M_h$ within the parameter range of \eqn{eq::c1max} is
shown in \fig{fig::mh}\,(b).

The choice of $\theta_t$ is motivated by the explicit result for
$c_{1}^{(0)}$ in \eqn{eq::c1lo} (see also
\eqn{eq::maxterm}), where the prefactor of the last term in the
first line becomes maximal for $\theta_t = \pi/4$. The expression in
brackets vanishes for $\mstop{1}=\mstop{2}$. However, in the limit
$\mstop{1}\ll\mstop{2}$ a term enhanced by $\mstop{2}^2/\mstop{1}^2$
survives which can dominate the result for $c_1^{(0)}$. This is shown in
Fig.~\ref{fig::C1}\,(b) where $C_1$ and $C_1^{\rm SM}$ are plotted as a
function of $\mstop{2}$.  One observes a rather strong variation of the
one-loop result of $C_1$. It even changes sign close to
$\mstop{2}\approx900$~GeV, where the \susy{} and the \sm{} contributions
cancel each other.  A similar behavior is observed at \nlo{}, where
$C_1$ vanishes for $\mstop{2}\approx850$~GeV.

\subsection{Hadronic decay rate}
For our numerical analysis we neglect all bottom and sbottom effects.  In
particular, the direct coupling of a Higgs boson to bottom quarks is not
contained in our formulae.  In this approximation the \lo{} result for
the hadronic decay of a light Higgs boson is determined through the
$\higgs\to gg$ amplitude shown in \fig{fig::dias}\,(a)--(c). At
higher orders, also multi-particle final states contribute, such as
$ggg$, $gq\bar q$, $ggq\bar q$, etc. ($q\neq t$). We write
\begin{equation}
\begin{split}
\Gamma(\higgs\to \mbox{hadrons}) \equiv
\Gamma^\higgs_g &= F_0\cdot\left(\frac{C_1}{c_1^{(0)}}\right)^2
\left(1 + \delta_{\rm PS}\right) =
F_0\cdot\left(\frac{\alpha_s}{3\pi}\right)^2\,\left( 1 +
  \delta \right)\,,
\label{eq::gammag}
\end{split}
\end{equation}
with
\begin{equation}
\begin{split}
F_0 &= 
\frac{M_\higgs^3\sqrt{2}\,G_{\rm F}}{8\pi}
\bigg|
g_t^\higgs\,A(\tau_t)
+ \frac{\mtop^2}{2\mstop{1}^2}\,g_{t,11}^\higgs\, \tilde{A}(\tau_1)
+ \frac{\mtop^2}{2\mstop{2}^2}\,g_{t,22}^\higgs\, \tilde{A}(\tau_2)
\bigg|^2\,,\\
\tau_t &= \frac{4\mtop^2}{M_\higgs^2}\,,\qquad
\tau_{i} = \frac{4\mstop{i}^2}{M_\higgs^2}\,,\quad i=1,2\,.
\label{eq::siglo}
\end{split}
\end{equation}
The second equality in \eqn{eq::gammag} illustrates our approach: the
exact leading order result proportional to $F_0$ is factored out, and the
corrections are treated in the effective-theory approach of
\eqn{eq::efflag}.  The quantity $\delta_{\rm PS}$ contains the real and
virtual corrections associated with the operator ${\cal O}_1$.  The
third equality in \eqn{eq::gammag} is obtained by expanding the ratio
$C_1/c_1^{(0)}$ in terms of $\alpha_s$. The coupling constants
$g_t^\higgs$ and $g_{t,ij}^\higgs$ in \eqn{eq::siglo} are defined in
Eqs.\,(\ref{eq::tophiggs})--(\ref{eq::stophiggs-1}), and
\begin{equation}
\begin{split}
A(t) &= \frac{3}{2}t\,\left[ 1 + (1-t)\,f(t) \right]\,,\qquad
\tilde{A}(t) = -\frac{3}{4}\,t\,\left[ 1-  t\,f(t) \right]\,,
\end{split}
\end{equation}
where
\begin{equation}
\begin{split}
f(t) &= \left\{
\begin{array}{ll}
\arcsin^2\left(\frac{1}{\sqrt{t}}\right)\,,&\quad t\geq 1\,,\\
-\frac{1}{4}\left[ \ln\frac{1+\sqrt{1-t}}{1-\sqrt{1-t}} - i\pi
  \right]^2\,,& \quad t < 1\,.
\end{array}
\right.
\end{split}
\end{equation}
For completeness, we remark that the limits for $t\to\infty$ are given
by
\begin{equation}
\begin{split}
\lim_{t\to\infty} A(t) &= 1\,,\qquad
\lim_{t\to\infty} \tilde{A}(t) = \frac{1}{4}\,,
\end{split}
\end{equation}
and thus,
\begin{equation}
\begin{split}
F_0 &\to \frac{M_\higgs^3\sqrt{2}\,G_{\rm
    F}}{8\pi}\,|c_1^{(0)}|^2\qquad
\mbox{for } \mtop,\mstop{},\mgluino\gg M_\higgs\,,
\end{split}
\end{equation}
where $c_1^{(0)}$ is given in \eqn{eq::c1lo}.
The quantity $\delta$ is expanded in terms of $\alpha_s$ as follows:
\begin{equation}
\begin{split}
\delta &= \api\delta^{(1)} + \left(\api\right)^2\delta^{(2)} +
\order{\alpha_s^3}\,,
\end{split}
\end{equation}
and similarly for $\delta_{\rm PS}$.
The relation between $\delta_{\rm PS}$ and $\delta$ is given by
\begin{equation}
\begin{split}
\delta^{(1)} &= \delta^{(1)}_{\rm PS} +
2\frac{c_1^{(1)}}{c_1^{(0)}}\,,\qquad
\delta^{(2)} = \delta^{(2)}_{\rm PS} +
 2\frac{c_1^{(1)}}{c_1^{(0)}}\delta^{(1)}_{\rm PS}
+2\frac{c_1^{(2)}}{c_1^{(0)}} + \left(\frac{c_1^{(1)}}{c_1^{(0)}}\right)^2
\,,\qquad
\label{eq::deldelps}
\end{split}
\end{equation}
with~\cite{\gghnloeff,Chetyrkin:1997iv}
\begin{equation}
\begin{split}
\delta^{(1)}_{\rm PS} &= \frac{73}{4} - \frac{7}{6}\,n_l + \left(
\frac{11}{2} - \frac{1}{3}\,n_l\right)\,\ln\frac{\muR^2}{M_\higgs^2}\,.
\\ \delta^{(2)}_{\rm PS} &=
\frac{37631}{96}-\frac{363}{8}\zeta(2)-\frac{495}{8}\zeta(3)
+\frac{2817}{16}\ln\frac{\muR^2}{M_\higgs^2}+\frac{363}{16}
\ln^2\frac{\muR^2}{M_\higgs^2}
\\& +n_l\left(-\frac{7189}{144}+\frac{11}{2}\zeta(2)+\frac{5}{4}\zeta(3)
-\frac{263}{12}\ln\frac{\muR^2}{M_\higgs^2}-\frac{11}{4}
\ln^2\frac{\muR^2}{M_\higgs^2}
\right) \\& +n_l^2\left(\frac{127}{108}-\frac{1}{6}\zeta(2)
+\frac{7}{12}\ln\frac{\muR^2}{M_\higgs^2}+\frac{1}{12}
\ln^2\frac{\muR^2}{M_\higgs^2}
\right)\,,
\label{eq::deltaps}
\end{split}
\end{equation}
where $n_l=5$ in our case.

$c_1^{(2)}$ is not known in the \mssm{}, thus only the \nlo{} result for
$\Gamma_g^\higgs$ can be calculated consistently up to now.  However,
along the lines of Ref.~\cite{Harlander:2003kf}, one can argue that the
numerical influence of $c_1^{(2)}$ is small at \nnlo{}, and that it is
justified to assume $c_1^{(2)} = c_1^{(2),{\rm SM}}$ as long as this
coefficient has not been computed in the \mssm{}. The motivation behind
this procedure is two-fold: On the one hand, one reduces the dependence
of the final result on the unphysical scales (this is more important for
the production rate to be discussed below). On the other hand, the
relative numerical influence of the coefficient $c_1^{(1)}$ through
\eqn{eq::deldelps} is more important at \nnlo{} than at \nlo{}.
Therefore, in the following we will give the numerical result for both 
\nlo{} and the estimated \nnlo{} (by setting $c_1^{(2)} = c_1^{(2),{\rm
    SM}}$). In order to indicate that this is not the full \nnlo{}
result, we denote it by \nnlo{}'.

Let us in the following discuss the numerical impact of $c_1^{(1)}$ to
the hadronic Higgs decay rate in the two scenarios discussed in
\sct{sec::c1res}.  In Fig.~\ref{fig::gam}\,(a) the decay rate is shown
as a function of $m_{1/2}$ where the thick dotted, dashed, and solid
line correspond to the \lo{}, \nlo{}, and \nnlo{}' prediction for the
{\abbrev SPS}\,1a scenario. For comparison, we show as thin lines the
corresponding \sm{} results.

Similarly, Fig.~\ref{fig::gam}\,(b) shows the decay rate as a function
of $\mstop{2}$ for the scenario of \eqn{eq::c1max}.  As expected, around
850~GeV $\Gamma_g^h$ is close to zero.  Furthermore, one observes a
screening of approximately 50\% and more for $\mstop{2} \ge 600$~GeV.
Note that according to \fig{fig::mh}\,(b), it is at $\mstop{2}\approx
600$\,GeV where $M_h$ assumes its maximal value; the region above
$\mstop{2}\approx 800$\,GeV, on the other hand, is experimentally
excluded due to the lack of a light Higgs signal at {\abbrev LEP}.

\FIGURE{
    \begin{tabular}{cc}
      \includegraphics[bb=110 265 465 560,width=18em]{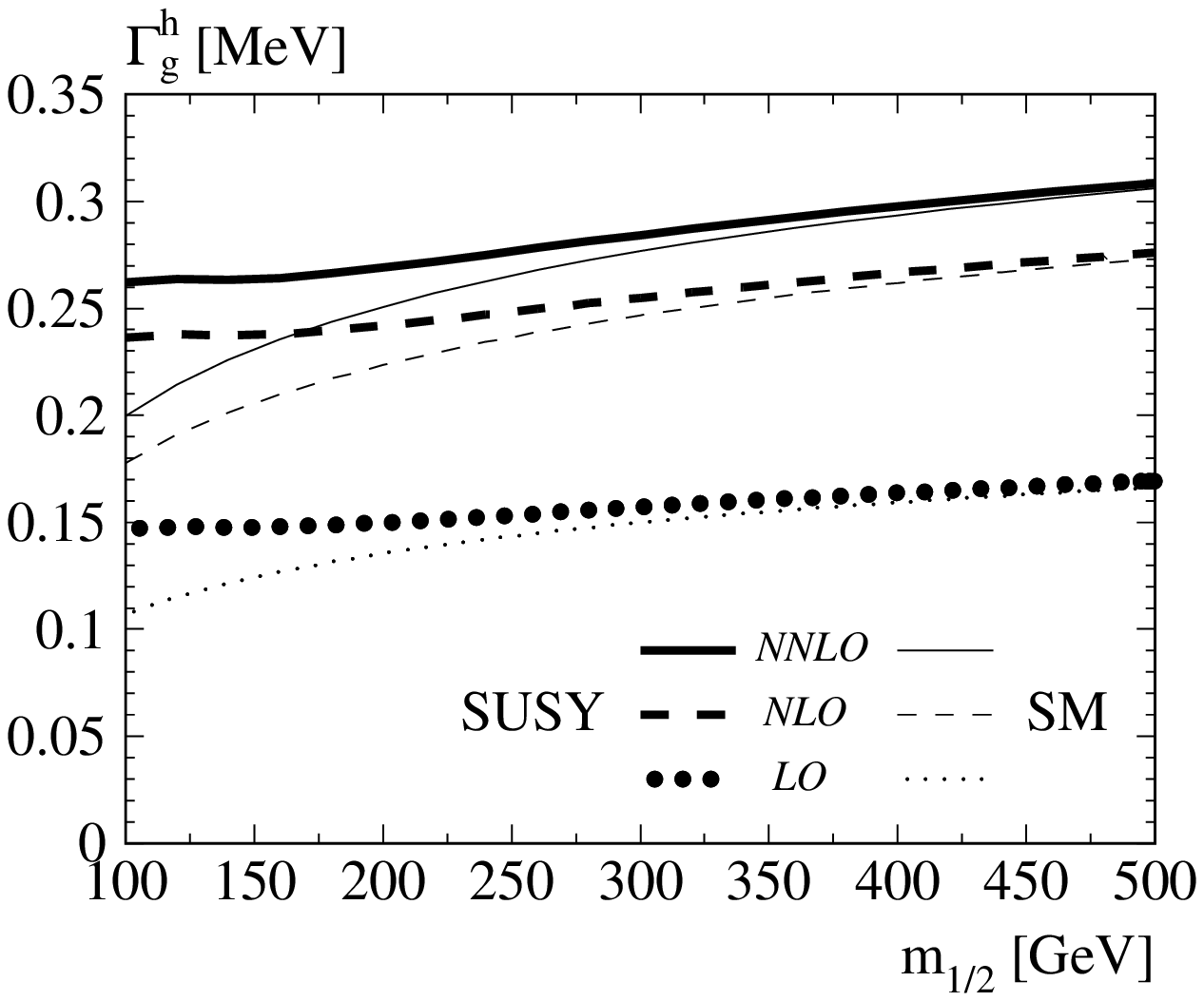} &
      \includegraphics[bb=110 265 465 560,width=18em]{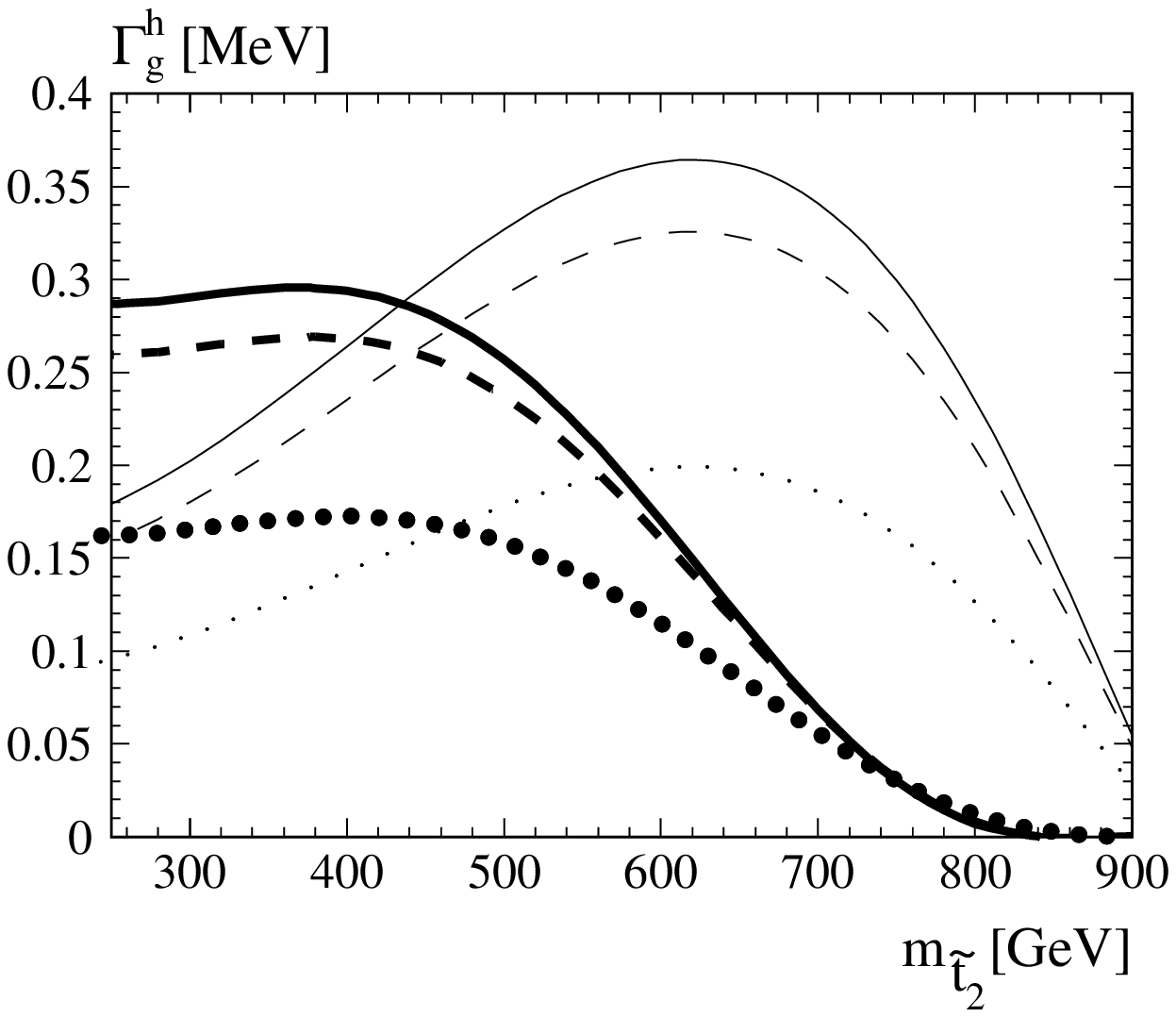}
      \\
      (a) & (b)
    \end{tabular}
    \parbox{14.cm}{
      \caption[]{\label{fig::gam}\sloppy
Hadronic decay rate for the light {\abbrev CP}-even Higgs boson
$h$ (a) along {\abbrev SPS}\,1a, and (b) for the scenario of
\eqn{eq::c1max}. The thick lines correspond to the \susy{} case,
the thin lines show the \sm{} results for comparison.
Solid: \nnlo{}'; dashed: \nlo{}; dotted: \lo{}.
    }}
}

\subsection{Hadronic production rate}
In the \sm{}, the total cross section as derived from the effective
Lagrangian of \eqn{eq::efflag} was shown to approximate the full result
to better than 3\% for $M_H<2\mtop$ {\it if} --- analogous to
\eqn{eq::gammag} --- the full top mass dependence at \lo{} is factored
out~\cite{Spira:1995rr,Spira:1997dg} (see also
Ref.~\cite{Harlander:2003xy}).
In this case, the hadronic
cross section $\sigma\subhk \equiv \sigma(pp\to \higgs+X)$ for Higgs
production can be written as
\begin{equation}
\begin{split}
  \sigma\subhk(z) &= \rho_0\,\left(\frac{C_1}{c_1^{(0)}}\right)^2\,
  \left[\Sigma^{(0)}\subhk(z) + \api\,\Sigma^{(1)}\subhk(z) +
  \left(\api\right)^2\,\Sigma^{(2)}\subhk(z) + \ldots\right]\\
  &= 
  \rho_0\,\left(\frac{\alpha_s}{3\pi}\right)^2\,
  \left[
    \Delta^{(0)}\subhk(z)
    + \api\,\Delta^{(1)}\subhk(z)
    + \left(\api\right)^2\,\Delta^{(2)}\subhk(z) + \ldots
    \right]\,,
  \label{eq::sigma}
\end{split}
\end{equation}
where
\begin{equation}
\begin{split}
\Sigma^{(n)}\subhk(z) &= \sum_{i,j\in\{q,\bar q,g\}}
  \int_z^1\dd x_1 \int_{z/x_1}^1
  \dd x_2\,\,
  \varphi_{i}(x_1)\,
  \varphi_{j}(x_2)\,
  \hat\Sigma^{(n)}_{ij}\left(\frac{z}{x_1x_2}\right)\,,\qquad
  z\equiv \frac{M_\higgs^2}{s}\,.
  \label{eq::sigmahk}
\end{split}
\end{equation}
$\varphi_{i}(x)$ is the density of parton $i$ inside the proton,
$M_\higgs$ is the Higgs boson mass, $s$ is the hadronic center-of-mass
(c.m.) energy, 
and\footnote{Note that in Eq.~(31) of Ref.~\cite{Harlander:2003bb} a factor
  $1/M_\phi^3$ is missing.}
\begin{equation}
\begin{split}
\rho_0 &= \frac{\pi^2}{8M_\higgs^3}\,F_0\,,
\label{eq::sigma0}
\end{split}
\end{equation}
with $F_0$ from \eqn{eq::siglo}.  In order to evaluate the \lo{},
\nlo{}, or the \nnlo{} cross section, the second line in \eqn{eq::sigma}
has to be truncated after the term $\Delta^{(0)}\subhk$,
$\Delta^{(1)}\subhk$, or $\Delta^{(2)}\subhk$, respectively.
Furthermore, the parton density functions ({\abbrev PDF}s) $\varphi_{i}$
in \eqn{eq::sigmahk} have to be used at the appropriate
order.\footnote{Only approximate \nnlo{} parton densities are currently
available; with the full \nnlo{} splitting functions being known
analytically now~\cite{Moch:2004pa,Vogt:2004mw}, this shortcoming is
expected to be eliminated in the near future.}  This results in
different values for $\Sigma^{(n)}\subhk(z)$ and
$\Delta^{(n)}\subhk(z)$, depending on the order that is being
considered. The same is true for $\alpha_s$ which has to be set in
accordance with the {\abbrev PDF} set.  Specifically, we adopt the
{\abbrev PDF} parameterizations of
Ref.~\cite{Martin:2002dr,Martin:2003tt} where $\alpha_s$ is given by
$0.1300$, $0.1165$, and $0.1153$ at \lo{}, \nlo{}, and \nnlo{},
respectively.

Note that at this point we choose a different value for $\alpha_s(M_Z)$
as the one defined in \eqn{eq::sminputs}. The latter enters the
evaluation of the low energy parameters through {\tt SoftSusy} or {\tt
SPheno}. This may be viewed as an inconsistency, but we find it more
natural to have the same set of \susy{} parameters at the various orders
of the calculation. Besides that, the spectrum calculators --- to our
knowledge --- do not provide control over the order of the evolution
equations, and the numerical effects of the value for $\alpha_s$ used
in \eqn{eq::sminputs} on the Higgs production cross section are small.

The \lo{} partonic result is
\begin{equation}
\begin{split}
\hat\Sigma_{ij}^{(0)}(x) &= \delta_{ig}\delta_{jg}\,\delta(1-x)\,.
\end{split}
\end{equation}

The \nlo{} quantity $\hat\Sigma_{ij}^{(1)}(x)$ can be derived from the
\sm{} expression of Refs.~\cite{Dawson:1990zj,Djouadi:1991tk}:
\begin{equation}
\begin{split}
  \hat\Sigma^{(1)}_{gg}(x) &= 6\zeta_2\, \delta(1-x) + 12 \left [
  \frac{\ln(1-x)}{1-x} \right ]_+ - 12x(-x+x^2+2)\ln(1-x) \\ & \quad
  -\frac{6 (x^2+1-x)^2}{1-x}\ln(x) -\frac{11}{2} (1-x)^3 \,, \\
  \hat\Sigma^{(1)}_{qg}(x) &= -\frac{2}{3} \left ( 1+(1-x)^2 \right )\ln
  \frac{x}{(1-x)^2}-1+2x -\frac{1}{3}x^2 \,, \\ 
  \hat\Sigma^{(1)}_{q\bar q}(x)
  &= \frac{32}{27}(1-x)^3 \,.
\end{split}
\end{equation}
The expression for $\hat\Sigma^{(2)}_{ij}$ is too long to be quoted
here. It can be extracted from Refs.~\cite{\gghnnlo}.  In analogy to the
discussion below \eqn{eq::deltaps}, we define an approximate \nnlo{}
result by setting $c_1^{(2)} = c_1^{(2),{\rm SM}}$ and denote it by
\nnlo{}'.

For convenience, we explicitly give the relation between the 
$\Delta^{(n)}\subhk$ and the $\Sigma^{(n)}\subhk$, $n=0,1,2$:
\begin{equation}
\begin{split}
\Delta\subhk^{(0)} &= \Sigma\subhk^{(0)}\,,\qquad
\Delta\subhk^{(1)} = \Sigma\subhk^{(1)} 
+ 2\frac{c_1^{(1)}}{c_1^{(0)}}\,\Sigma\subhk^{(0)}\,,\\
\Delta\subhk^{(2)} &= \Sigma\subhk^{(2)} 
+ 2\frac{c_1^{(1)}}{c_1^{(0)}}\,\Sigma\subhk^{(1)}
+ \left[\left(\frac{c_1^{(1)}}{c_1^{(0)}}\right)^2
  + 2\frac{c_1^{(2)}}{c_1^{(0)}}\right]\,\,\Sigma\subhk^{(0)}\,.
\label{eq::DelSig}
\end{split}
\end{equation}
The quantities $\Sigma\subhk^{(n)}$ are independent of the specific
model under consideration. A publicly available numerical program for
their evaluation is in preparation~\cite{ggh@nnlo}. 

For illustration of the numerical effects on the total Higgs production
cross section in gluon fusion, we consider again the two exemplary cases
of \sct{sec::c1res}.  The thick lines of \fig{fig::sps1a-ggh}\,(a) show
the \lo{} (dotted), \nlo{} (dashed), and \nnlo{}' (solid) cross
section for the {\abbrev SPS}\,1a scenario, compared to the \sm{}
prediction (thin lines). At $m_{1/2} = 100$\,GeV, the \mssm{} values are
about 34\% (32\%) larger than the \sm{} values at \nlo{}
(\nnlo{}'), but the difference decreases quite rapidly when $m_{1/2}$
increases.  However, this difference is mostly a \lo{} effect, as can be
seen from \fig{fig::sps1a-ggh}\,(b) which shows the \nlo{} and the
\nnlo{}' K-factor in the \mssm{} and the \sm{}. The difference of the
relative corrections in both cases is less than 5\%.

\fig{fig::c1max-ggh}\,(a), on the other hand, corresponds to the
scenario defined in \eqn{eq::c1max}.  As seen in \fig{fig::C1}\,(b),
$C_1$ vanishes for a certain value of $\mstop{2}$, and so does the
\nlo{} and the \nnlo{}' cross section. Note, however, that this
particular value is experimentally excluded because it corresponds to a
too low Higgs mass (see \fig{fig::mh}\,(b)).  Nevertheless, for
$\mstop{2}\approx 600$\,GeV, where $M_h$ is maximal, the cross section
is still significantly suppressed with respect to the \sm{}.  As the
K-factor in \susy{} tends to be a little smaller than in the \sm{}, this
suppression becomes even stronger when \qcd{} corrections are included.
For example, at $\mstop{2}\approx 600$\,GeV (or alternatively,
$|X_t|\approx 900$\,GeV), the ratio 
$\sigma^{\rm MSSM}/\sigma^{\rm SM}$ is 0.58 at
\lo{}, 0.52 at \nlo{}, and 0.48 at \nnlo{}'.

\FIGURE{
    \begin{tabular}{cc}
      \includegraphics[bb=110 265 465 560,width=18em]{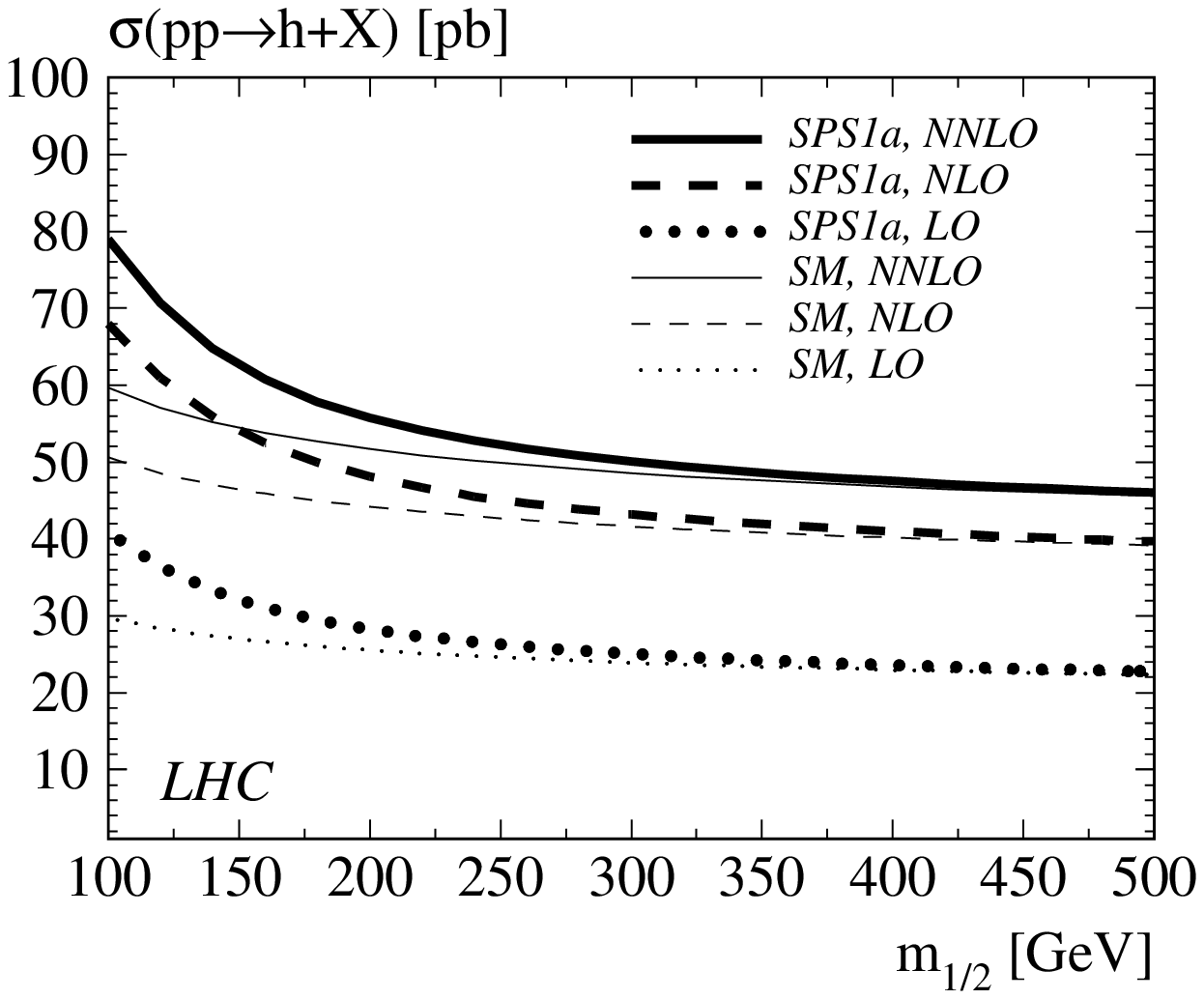} &
      \includegraphics[bb=110 265 465 560,width=18em]{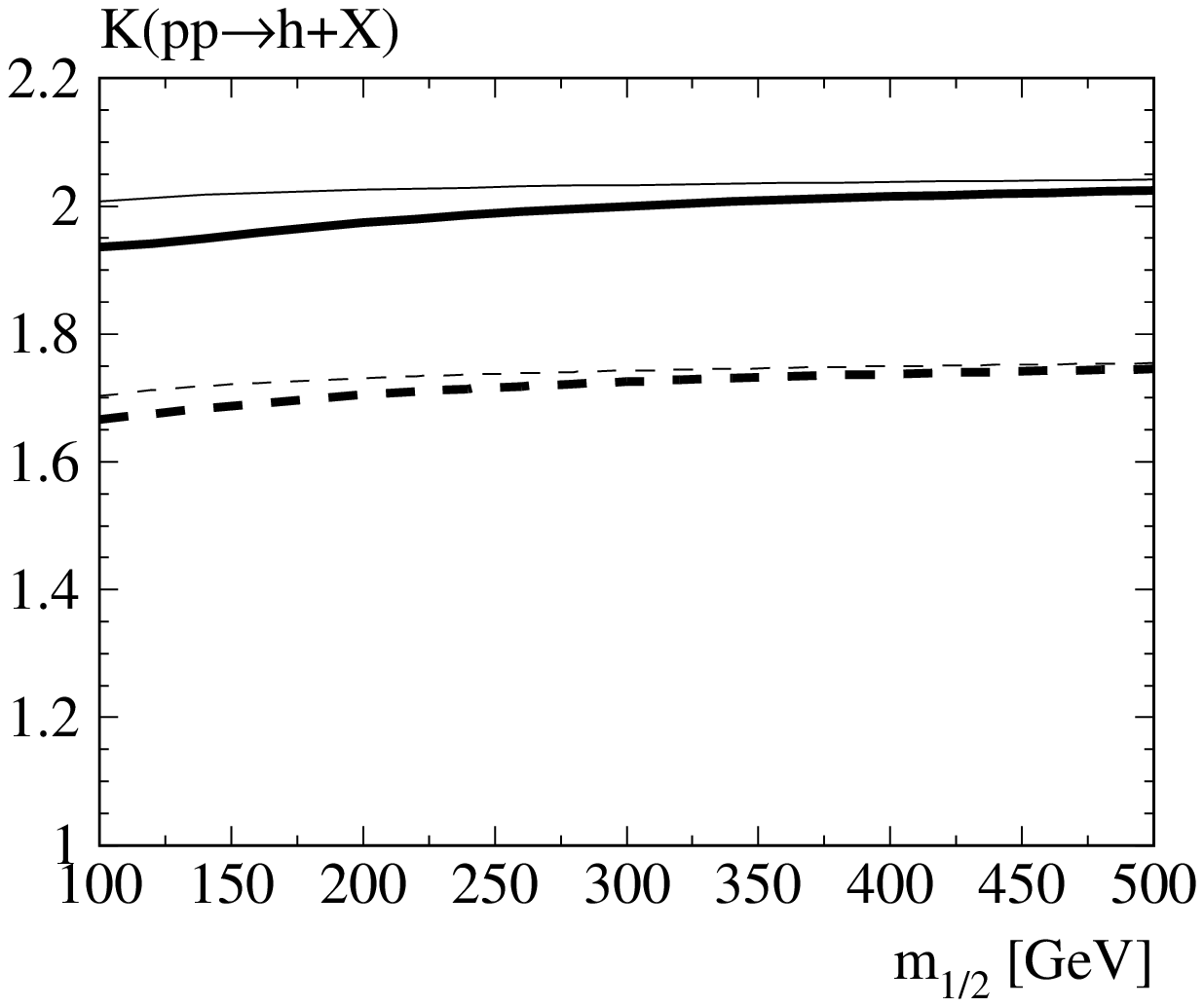}
      \\
      (a) & (b)
    \end{tabular}
      \parbox{14.cm}{
\caption[]{\label{fig::sps1a-ggh}\sloppy (a) Total cross section at
  \lo{} (dotted), \nlo{} (dashed), and \nnlo{}' (solid) along the slope
  of {\abbrev SPS}\,1a (thick lines). For comparison, also the \sm{}
  result is shown (thin lines). It depends on $m_{1/2}$ due to the
  variation of $M_h$ with this parameter.  (b) \nlo{} and \nnlo{}'
  K-factor for the {\abbrev SPS}\,1a scenario (thick lines) and for the
  \sm{} (thin lines).}}
}

\FIGURE{
    \begin{tabular}{cc}
      \includegraphics[bb=110 265 465 560,width=18em]{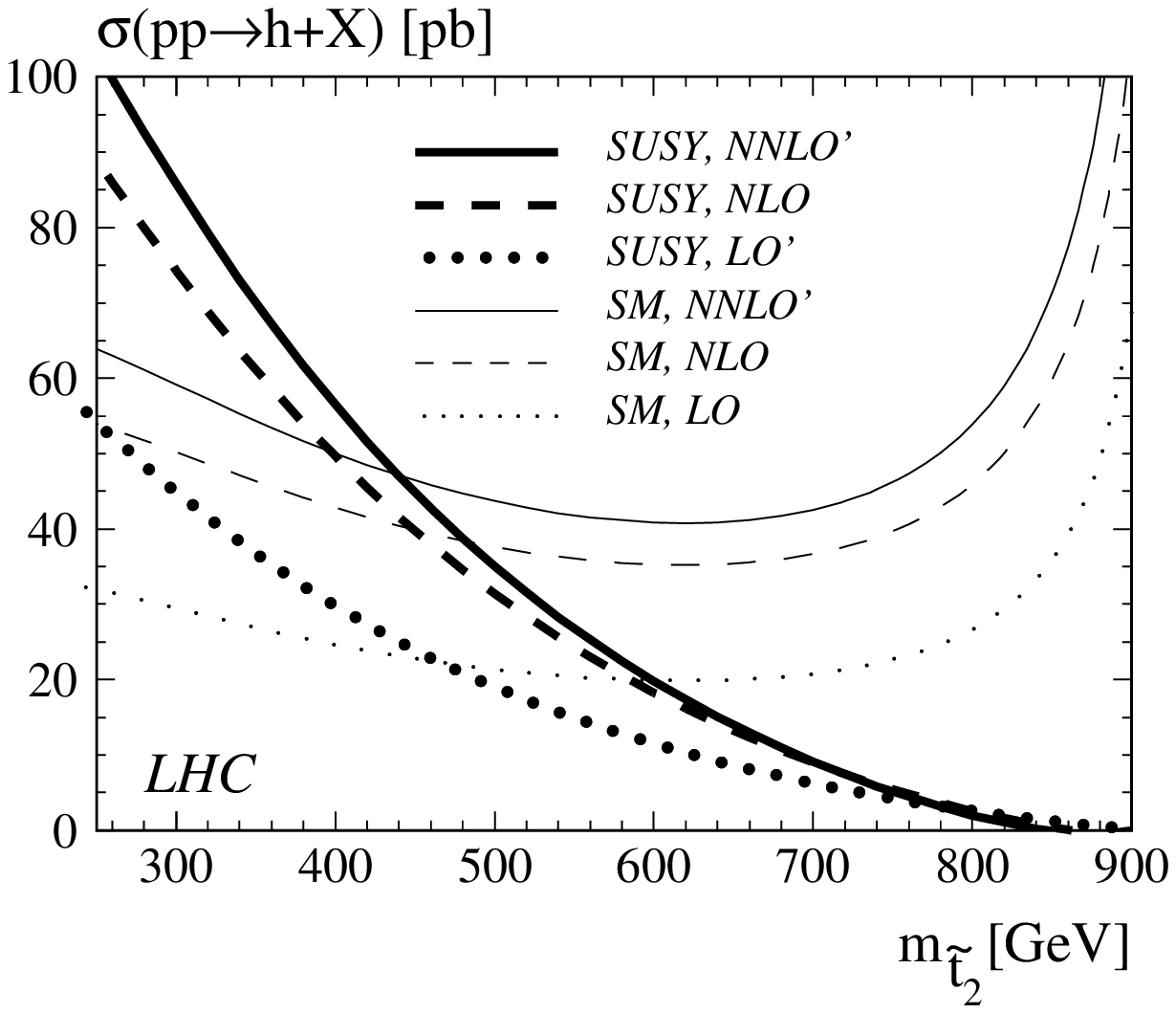} &
      \includegraphics[bb=110 265 465 560,width=18em]{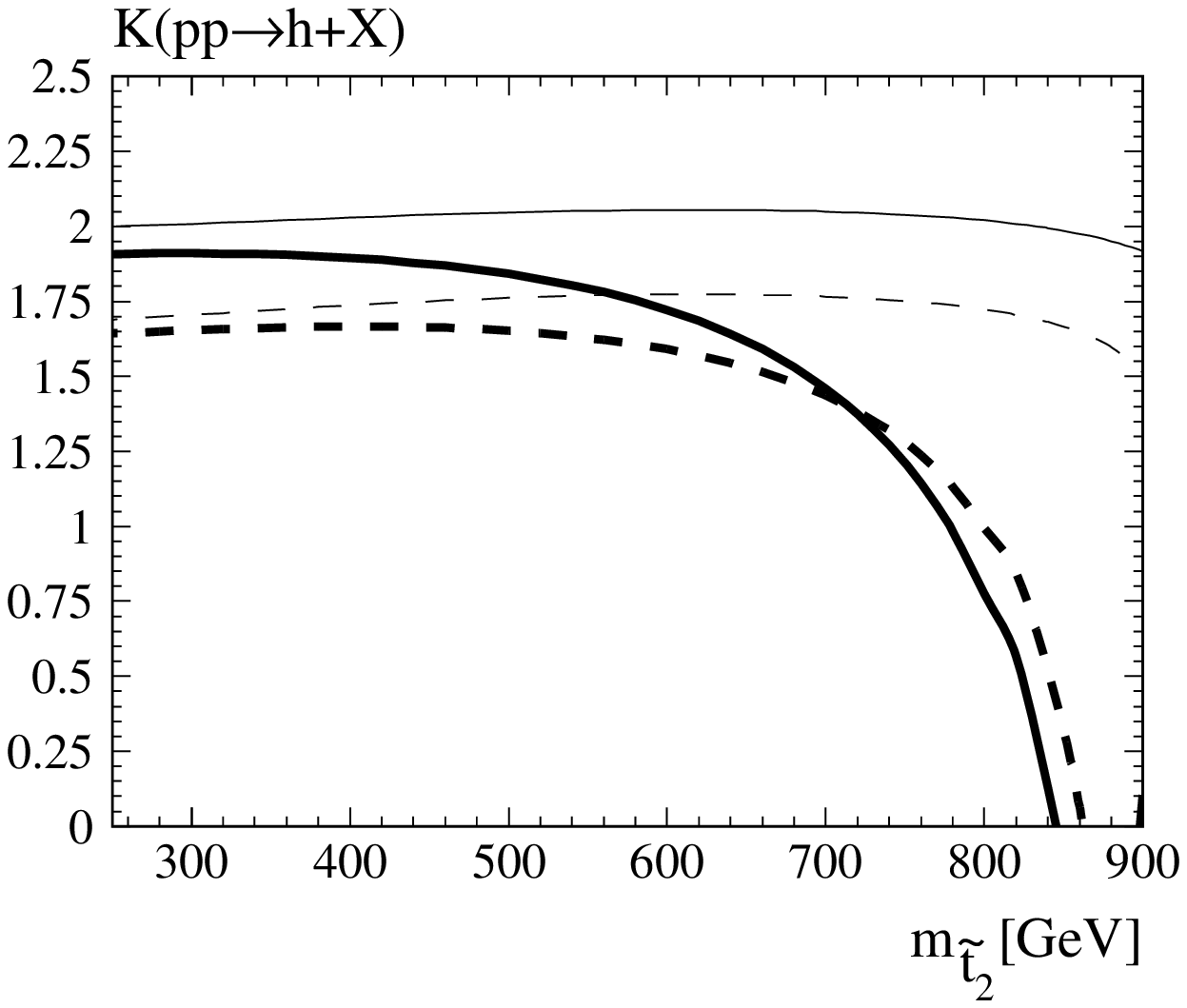}
      \\
      (a) & (b)
    \end{tabular}
    \parbox{14.cm}{
      \caption[]{\label{fig::c1max-ggh}\sloppy (a) Cross section and (b)
K-factor for the scenario defined in \eqn{eq::c1max} (thick lines) and
for the \sm{} (thin lines).  Dotted lines correspond to \lo{}, dashed
lines to \nlo{}, and solid lines to \nnlo{}'.}}
}

\subsection{Discussion}
In a model where the gluon-Higgs coupling is mediated predominantly by
heavy particles, it had already been observed that the radiative
corrections to the hadronic production and decay processes are not very
sensitive to the specifics of this
coupling~\cite{Dawson:1996xz,Harlander:2003kf}. This
is due to the fact that the radiative corrections are dominated by soft
gluon effects which do not resolve the gluon-Higgs
vertex~\cite{Harlander:2001is,Catani:2001ic,
Harlander:2002wh,Catani:2003zt}.\footnote{It is remarkable, however, that
the resummation of these soft-gluon effects gives only an effect of
about 6\%~\cite{Catani:2003zt} with respect to the fixed order \nnlo{}
result. This is usually interpreted as a sign of the stability of the
\nnlo{} prediction.}

Aside from this, for typical \mssm{} benchmark points, even the Wilson
coefficient of the effective gluon-Higgs coupling itself is numerically
rather close to its \sm{} value, both at \lo{} and \nlo{}. Only if at
least one of the scalar top quarks is relatively light
($\mstop{}\lesssim 400$\,GeV), a significant deviation from the \sm{}
result is observed. This is because the stop Yukawa coupling is
proportional to $\mtop^2$ rather than $\mstop{}^2$. In combination with
the loop amplitude of \fig{fig::dias}\,(b),(c), this leads to a
suppression factor $\mtop^2/\mstop{}^2$. In contrast to this, for the
top quark contribution there is a cancellation between the Yukawa
coupling $\sim \mtop$ and a factor $1/\mtop$ from the loop amplitude.

We pointed out that this suppression of the squark contribution may be
compensated by a large absolute value of the parameter $X_t = A_t -
\muSUSY{}\cot\beta$. According to \eqn{eq::maxterm}, this corresponds to
a stop mixing angle of the order of $\theta_t=\pi/4$, and a large mass
splitting between $\mstop{1}$ and $\mstop{2}$.  However, the value of
$X_t$ is crucial for the exact value of the light Higgs boson mass
$M_h$. This restricts $|X_t|$ to less than about 3\,TeV.  In
\eqn{eq::c1max}, we have chosen a set of low energy parameters which
fulfills this condition, but where the Wilson coefficient $C_1$ for the
gluon-Higgs interaction is very different from its \sm{} value (see
\fig{fig::C1}\,(b)) and leads to a strongly reduced production and decay
rate. Since also here the \qcd{} corrections tend to be smaller than in
the \sm{}, this cancellation effect of top and stop contributions is even
stronger when \qcd{} corrections are included.

\section{Conclusions}\label{sec::conclusions}
We have analytically calculated the \nlo{} \qcd{} contribution to the
effective gluon-Higgs coupling in the \mssm{} due to the scalar partners
of the top quark. Scalar bottom effects are generally suppressed by
$m_b^2/m_{\tilde b}^2$ or $m_b\muSUSY/m_{\tilde b}^2$ and have been
neglected. The calculation involves Feynman diagrams with three massive
particles (gluino, top quark, stop quark) which leads to very long
analytic expressions for the final result.  Therefore, we make it
available in the form of a {\tt Fortran} routine, described in
\appx{sec::evalcsusy}.

The results for the effective coupling were used to evaluate the
hadronic Higgs decay rate and the production cross section through
\nlo{} in \qcd{}, and to derive a \nnlo{} estimate of these quantities.
The \qcd{} corrections in the \mssm{} tend to be a bit smaller than in
the \sm{}. However, this effect is in general below 5\%.  In regions of
the \mssm{} parameter space where the Higgs coupling to gluons is
particularly small due to a cancellation between the quark and the squark
contribution, the reduced K-factor amplifies this effect.  Nevertheless,
even here the K-factor in the \sm{} provides a fairly accurate
approximation to the \mssm{} value.

We conclude by noting that the methods of our calculation should be
immediately applicable to the photonic production and decay rate of an
\mssm{} Higgs boson, as well as to pseudo-scalar Higgs
production. Inclusion of sbottom effects is also possible, but requires
a careful treatment of the bottom threshold in the Feynman diagrams.

\paragraph{Acknowledgments.}
The authors are grateful to M.~Faisst, S.~Heinemeyer, and S.~Kraml for
useful discussions. RVH is supported by {\it Deutsche
Forschungsgemeinschaft}, contract HA~2990/2-1.

\newpage
\begin{appendix}

\section{Feynman rules}\label{sec::frules}
In this appendix, we collect the Feynman rules that have been used in
our calculation of the \nlo{} coefficient function $C_1$. The notation
follows closely Ref.~\cite{Kraml-Eberl}.

\subsection{Definitions}\label{sec::defs}

In the following, $p$, $k$, and $p_n$ ($n=1,2,3$) denote incoming
four-momenta; the various indices have the following meaning:
\begin{center}
\begin{tabular}{ll}
  $r,s,t,u$ &: color triplet indices\\
  $a,b,c$ &: color octet indices\\
  $\mu,\nu,\rho$ &: Lorentz indices\\
  $i,j,k,l$ &: squark mass eigenstate indices\\
  $A,B$ &: flavor indices
\end{tabular}
\end{center}
Furthermore, we introduce
\begin{equation}
\begin{split}
&[T^a,T^b] = if^{abc}T^c\,,\qquad 
\{T^a,T^b\} = \frac{1}{n_c}\delta^{ab} + d^{abc}T^c\,,\\
&P_L = \frac{1-\gamma_5}{2}\,,\qquad
P_R = \frac{1+\gamma_5}{2}\,,\\
&{\cal R}^q = \left(
\begin{array}{cc}
  {\cal R}^q_{11} & {\cal R}^q_{12} \\
  {\cal R}^q_{21} & {\cal R}^q_{22}
\end{array}
\right)
=
\left(
\begin{array}{cc}
  \cos\theta_q& \sin\theta_q \\
  -\sin\theta_q & \cos\theta_q
\end{array}
\right)\,,\\
&{\cal S}^q = 
\left(
\begin{array}{cc}
  {\cal S}^q_{11} & {\cal S}^q_{12} \\
  {\cal S}^q_{21} & {\cal S}^q_{22}
\end{array}
\right)
=\left(
\begin{array}{cc}
  \cos2\theta_{q}& -\sin2\theta_{q} \\
  -\sin2\theta_{q} & -\cos2\theta_{q}
\end{array}
\right)
\,.
\end{split}
\end{equation}
$\theta_q\in [0,\frac{\pi}{2})$ is the squark mixing angle defined through
\begin{equation}
\begin{split}
\sin 2\theta_q = \frac{2m_qa_q}{m_{\tilde q_1}^2 - m_{\tilde q_2}^2}\,,\qquad
\cos 2\theta_q = \frac{m_{\tilde q_L}^2 - m_{\tilde
  q_R}^2}{m_{\tilde q_1}^2 - m_{\tilde q_2}^2}\,,
\label{eq::thetaqdef}
\end{split}
\end{equation}
where, by definition, we assume $m_{\tilde q_1}\leq m_{\tilde q_2}$, and
\begin{equation}
\begin{split}
X_q = A_q - \muSUSY\cdot\left\{\begin{array}{c}
  \cot\beta\,,\quad \mbox{for}\quad q\in\{u,c,t\}\\
  \tan\beta\,,\quad \mbox{for}\quad q\in\{d,s,b\}
 \end{array}
\right.
\,.
\end{split}
\end{equation}
$A_q$ and $\muSUSY$ are soft \susy{} breaking parameters (see
Ref.\cite{Martin:1997ns}, for example).

\subsection{Feynman rules used in this calculation}\label{app::frules}
\begin{tabular}{ll}
\raisebox{-4.5em}{
  \includegraphics[bb=120 550 310 725,width=10em]{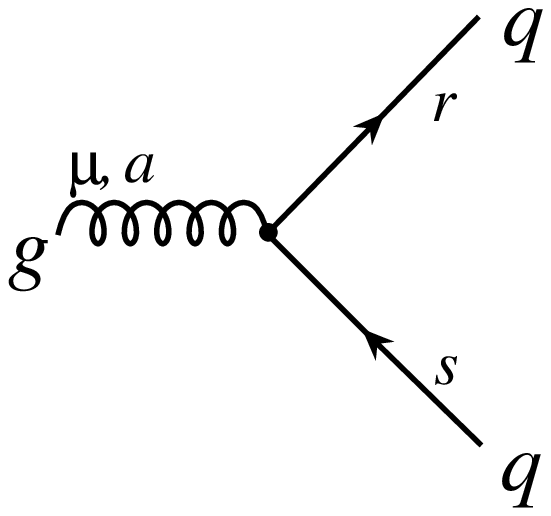}} & 
$\ds ig_sT_{rs}^a\gamma^\mu$
\\
\raisebox{-4.5em}{
\includegraphics[bb=120 550 310 725,width=10em]{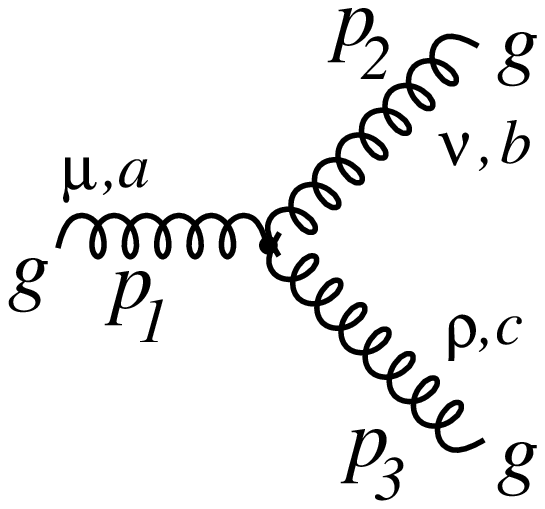}} & 
$ g_s f^{abc}\left[ (p_1-p_2)^\rho g^{\mu\nu} + (p_2-p_3)^\mu g^{\nu\rho}
  + (p_3-p_1)^\nu g^{\mu\rho} \right]$
\\
\raisebox{-4.5em}{
\includegraphics[bb=120 550 310 725,width=10em]{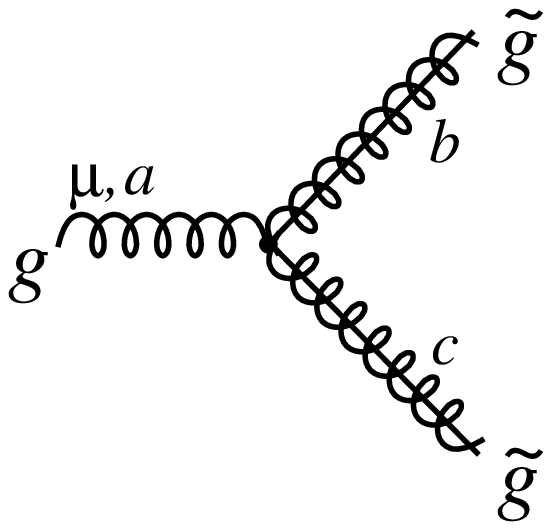}} & 
$g_s f^{abc}\gamma^\mu$
\\
\raisebox{-4.5em}{
\includegraphics[bb=120 550 310 725,width=10em]{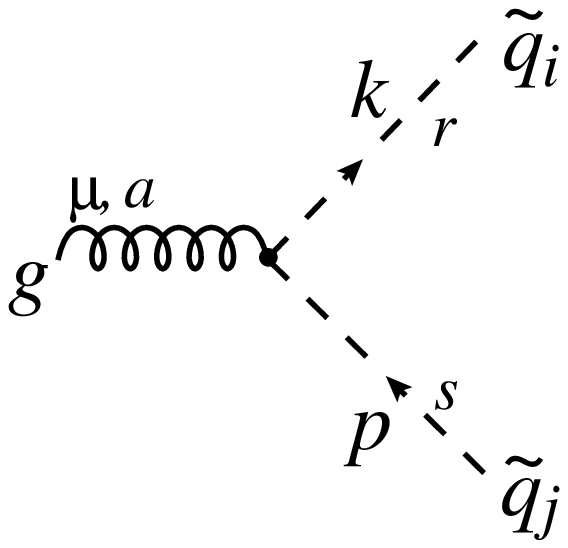}} & 
$\ds ig_sT^a_{rs}(p-k)^\mu\delta_{ij}$
\\
\raisebox{-4.5em}{
\includegraphics[bb=120 550 310 725,width=10em]{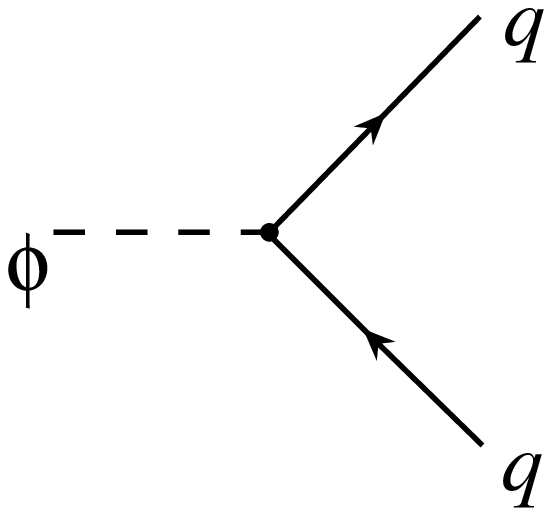}} & 
$\ds i \frac{m_q}{v}\,g_q^\phi$
\\
\raisebox{-4.5em}{
\includegraphics[bb=120 550 310 725,width=10em]{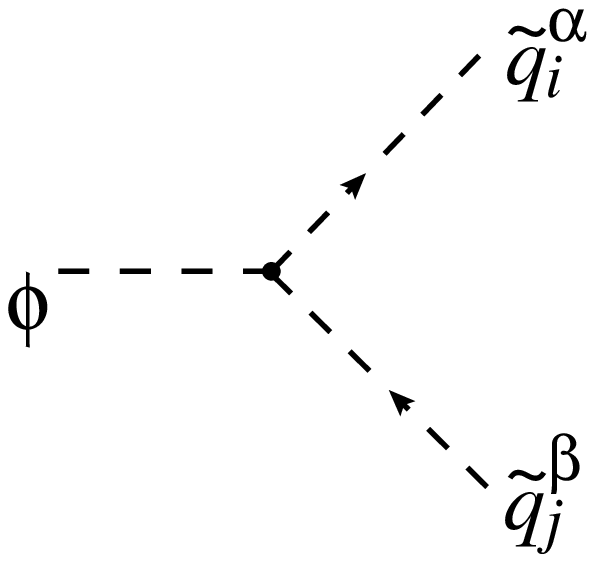}} & 
$\ds i \frac{m_q^2}{v}\,g_{q,ij}^{\phi}$
\end{tabular}

\begin{tabular}{ll}
\raisebox{-4.5em}{
\includegraphics[bb=120 550 310 725,width=10em]{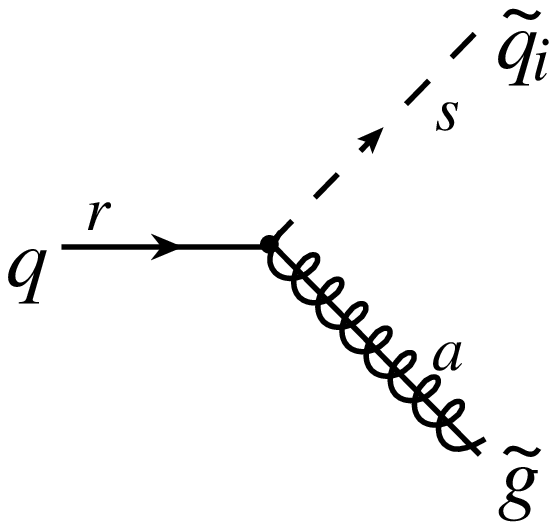}} & 
$ig_sT^a_{rs}\sqrt{2}\left(
{\cal R}_{i1}^q\,P_L - {\cal R}_{i2}^q\,P_R\right)$
\\
\raisebox{-4.5em}{
\includegraphics[bb=120 550 310 725,width=10em]{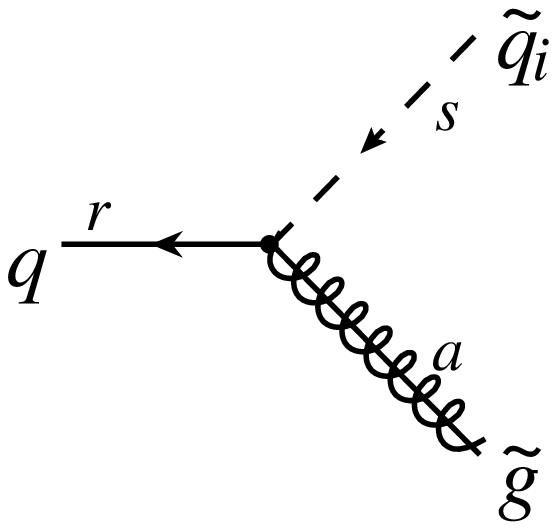}} & 
$ig_sT^a_{rs}\sqrt{2}\left(
{\cal R}_{i1}^q\,P_R - {\cal R}_{i2}^q\,P_L\right)$
\\
\raisebox{-4.5em}{
\includegraphics[bb=120 550 310 725,width=10em]{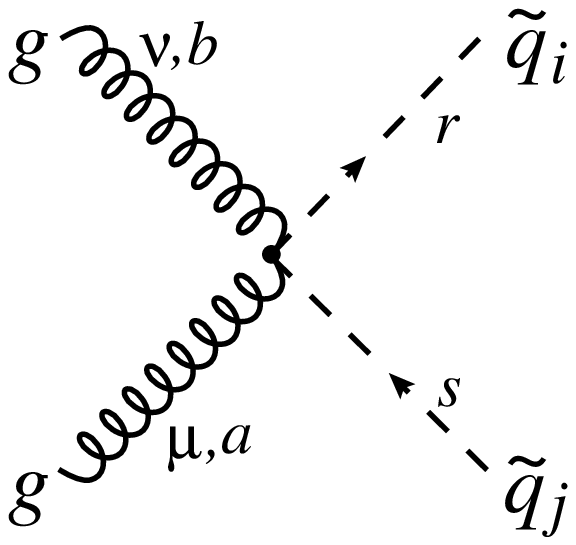}} &
$-i g_s^2\left( \frac{1}{3}\delta_{ab}\delta_{rs} +
d_{abc}T^c_{rs}\right) g_{\mu\nu}\,\delta_{ij}$ \\ 
\raisebox{-4.5em}{
\includegraphics[bb=120 550 310 725,width=10em]{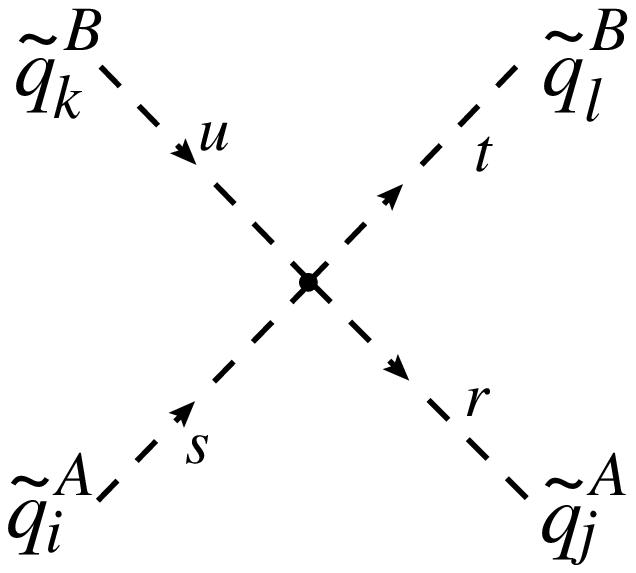}} & 
$ig_s^2\left[
  T^a_{rs}T^a_{tu}{\cal S}_{ij}^{A}{\cal S}_{kl}^B
  + T^a_{ru}T^a_{ts}{\cal S}_{il}^{A}{\cal
    S}_{kj}^A\delta_{AB}\right]$\,,\qquad 
(${\cal S}^A \equiv {\cal S}^{q^A}$)
\\
\end{tabular}

We only give the values for the couplings to the light Higgs here; the
couplings to the heavy Higgs can be obtained from the latter through the
replacement $\alpha \to \alpha + \pi/2$.
The top-Higgs coupling is
\begin{equation}
\begin{split}
g_t^h &=\frac{\cos\alpha}{\sin\beta}\,,
\label{eq::tophiggs}
\end{split}
\end{equation}
and the stop-Higgs couplings are
\begin{equation}
\begin{split}
g_{t,ij}^\phi &=g_{t,ij}^{\phi,\rm EW}
+ g_{t,ij}^{\phi,\mu} + g_{t,ij}^{\phi,\alpha}\,,
\label{eq::stophiggs}
\end{split}
\end{equation}
with
\begin{equation}
\begin{split}
g_{t,11}^{h,{\rm EW}} &= \cew{1}\,\cos^2\theta_t +
\cew{2}\,\sin^2\theta_t\,,\\
g_{t,22}^{h,{\rm EW}} &= \cew{1}\,\sin^2\theta_t + \cew{2}\,\cos^2\theta_t\,,\\
g_{t,12}^{h,{\rm EW}} &= 
g_{t,21}^{h,{\rm EW}} = 
\frac{1}{2}(\cew{2}-\cew{1})\,\sin 2\theta_t\,,\\
g_{t,11}^{h,\mu} &= - g_{22}^{h,\mu} = 
\frac{\muSUSY}{\mtop}\,\frac{\cos(\alpha-\beta)}{\sin^2\beta}
\sin2\theta_t\,,\\
g_{t,12}^{h,\mu} &= 
g_{t,21}^{h,\mu} = 
\frac{\muSUSY}{\mtop}\,\frac{\cos(\alpha-\beta)}{\sin^2\beta}
\cos2\theta_t\,,\\
g_{t,11}^{h,\alpha} &= \frac{\cos\alpha}{\sin\beta}\left[
  2 + \frac{\mstop{1}^2 - \mstop{2}^2}{2\mtop^2}\,
\sin^22\theta_t\,\right],\\
g_{t,22}^{h,\alpha} &= \frac{\cos\alpha}{\sin\beta}\left[
  2 - \frac{\mstop{1}^2 - \mstop{2}^2}{2\mtop^2}\,
\sin^22\theta_t\,\right],\\
g_{t,12}^{h,\alpha} &= 
g_{t,21}^{h,\alpha} = 
\frac{\cos\alpha}{\sin\beta}\,
  \frac{\mstop{1}^2 - \mstop{2}^2}{2\mtop^2}\,
\sin2\theta_t\cos2\theta_t\,,\\
\label{eq::stophiggs-1}
\end{split}
\end{equation}
where
\begin{equation}
\begin{split}
\cew{1} &= -\frac{M_Z^2}{\mtop^2}\left( 1 - \frac{4}{3}\sin^2\theta_W
\right)\,\sin(\alpha+\beta)\\ 
\cew{2} &= -
\frac{M_Z^2}{\mtop^2}\,\frac{4}{3}\,\sin^2\theta_W\sin(\alpha+\beta)\,,\\
\sin\theta_W &= \sqrt{1 - \frac{M_W^2}{M_Z^2}}\,,
\label{eq::cew}
\end{split}
\end{equation}
and 
\begin{equation}
\begin{split}
v = \frac{2M_W}{g} = \frac{1}{\sqrt{\sqrt{2}\gfermi}} =
\sqrt{v_1^2+v_2^2} \approx 246\,{\rm GeV}\,,
\end{split}
\end{equation}
with $v_1$, $v_2$ the vacuum expectation values of the two Higgs doublets.
In \eqn{eq::stophiggs-1} we have already expressed the trilinear couplings
of the soft \susy{} breaking terms through independent parameters:
\begin{equation}
\begin{split}
A_t &= \frac{\mstop{1}^2 - \mstop{2}^2}{2\mtop}\sin2\theta_t +
\muSUSY\cot\beta\,.
\end{split}
\end{equation}
The electroweak radiative corrections to this formula need not be
considered here.

\section{Renormalization and decoupling constants}\label{sec::renorm}

In order to arrive at a finite \nlo{} result, the parameters appearing
in the \lo{} coefficient function given in \eqn{eq::c1lo} have to be
renormalized. This includes the strong coupling constant $\alpha_s$, the
top quark mass $\mtop$, the top squark masses $\mstop{i}$, and the
mixing angle $\theta_t$, whereas the angles $\alpha$, $\beta$ and
$\theta_W$ are not renormalized, because we consider \qcd{} corrections
only.  

First, the top quark and the \susy{} partners are decoupled from the
bare coupling constant through the relation
\begin{equation}
\begin{split}
  \alpha_s^\bare  &=  \left(\zeta_g^\bare\right)^2 \tilde{\alpha}_s^\bare
  \,,
  \label{eq::asdec}
\end{split}
\end{equation}
where $\tilde\alpha_s^\bare$ and $\alpha_s^\bare$ denote the bare
couplings in the full theory and in five-flavor \qcd{}, respectively.
In the $\drbar$ scheme, we find
\begin{equation}
\begin{split}
(\zeta_g^\bare)^2 &= 1 - \api\left(\frac{1}{\ep}
  \left[\frac{1}{6}C_A + \frac{1}{2}T\right]
  + L(\ep)\right) +
  \order{\alpha_s^2}\,,
\end{split}
\end{equation}
where $C_A=3$, $T=1/2$ and
\begin{equation}
\begin{split}
  L(\ep) &=
  \frac{1}{12}\left[ 2C_A\,L_{\tilde{g}} +  
  T (L_{\tilde{t}1} + L_{\tilde{t}2} + 4 \, L_t) \right]
  \\&\quad+
  \frac{\ep}{12}\left[
    C_A\,L_{\tilde{g}}^2 
    + \frac{1}{2}T\left(L_{\tilde{t}1}^2 + L_{\tilde{t}2}^2\right) 
    + 2T\, L_t^2 + \left(C_A+3 T\right)\,\zeta_2 \right]\,,\\
\Ltop &= \ln\frac{\muR^2}{\mtop^2}\,,\qquad
\Lstop{i} = \ln\frac{\muR^2}{\mstop{i}^2}\,,\qquad
\Lgluino = \ln\frac{\muR^2}{\mgluino}\,.
    \label{eq::l}
\end{split}
\end{equation}
$\alpha_s^\bare$ is then renormalized through
\begin{equation}
\begin{split}
  \frac{\alpha_s^\bare}{\alpha_s^\drbar} &= 1 +
  \frac{\alpha_s}{\pi}\frac{1}{\epsilon}
  \left[ - \frac{11}{12} C_A 
    + \frac{1}{3} T n_l
    \right]
  \,,
\label{eq::as}
\end{split}
\end{equation}
where $\alpha_s^\drbar$ denotes the $\drbar$ expression for the strong
coupling constant in \qcd{} with $n_l=5$ active flavors. 
 We will comment on the transformation to the more
familiar $\msbar$ scheme below.

For the quark and squark masses, we adopt the on-shell scheme, where
they are defined as the real part of the pole of the corresponding
propagator.  Furthermore, we define the renormalized squark mixing angle
by requiring that the non-diagonal two-point function
$\langle \tilde{t_1}\tilde{t_2} \rangle$ vanishes at a certain momentum
transfer $q_0$; i.e., the two squarks propagate independently from each
other at the scale $q_0$. In practice, $q_0$ is chosen to be of the
order of the squark masses.  The counter terms can be found in
Ref.~\cite{Djouadi:1998sq}, for example.  For convenience, we list them
explicitely in our notation.

In \dred{}, the relation between the bare and the pole top quark mass 
reads
\begin{eqnarray}
  \frac{\mtop^\bare}{\mtop} &=&
  1 + C_F\frac{\alpha_s}{\pi}\Bigg\{
  -\frac{1}{2\epsilon} 
  -\frac{5}{4}
  -\frac{3}{4} \Ltop
  -\frac{\mgluino^2}{4\mtop^2}\left(1+\Lgluino\right)
  +\sum_{i=1}^{2}\Bigg[
    \frac{\mstop{i}^2}{8\mtop^2}\left(1+\Lstop{i}\right)
  \nonumber\\&&\mbox{}
    +\frac{1}{8}\left(1+\frac{\mgluino^2}{\mtop^2}
      -\frac{\mstop{i}^2}{\mtop^2}
      + 2\,(-1)^i\frac{\mgluino}{\mtop}
      \sin 2\theta_t\right) B_0^{\rm fin}(\mtop^2,\mgluino,\mstop{i})
  \Bigg]
  \Bigg\}
  \,,
\label{eq::mtpole}
\end{eqnarray}
where $C_F=4/3$.
The only difference between \dred{} and Dimensional Regularization (\dreg{})
comes from the gluon-exchange diagram which changes the constant ``$5/4$''
into ``$1$'' in the case of \dreg{}.
The relation for the top squark mass $\mstop{1}$ is given by
\begin{equation}
\begin{split}
  \frac{\mstop{1}^\bare}{\mstop{1}} &=
  1 + C_F\frac{\alpha_s}{\pi}\Bigg\{
  \frac{1}{8\mstop{1}^2\, \ep}\left[
    4\mgluino\mtop \sin 2\theta_t 
      - 4\mgluino^2 - 4\mtop^2 + (\mstop{2}^2-\mstop{1}^2)\sin^2 2\theta_t
  \right]
  \\&
  -\frac{3}{4} - \frac{\sin^2 2\theta_t}{8}
  -\left(\frac{1}{4}+\frac{\sin^2 2\theta_t}{8}\right) \Lstop{1}
  -\frac{\mgluino^2} {4\mstop{1}^2}\left(1+\Lgluino\right)
  -\frac{\mtop^2}    {4\mstop{1}^2}\left(1+\Ltop\right)
  \\&
  +\frac{\mstop{2}^2\sin^2 2\theta_t}
                     {8\mstop{1}^2}\left(1+\Lstop{2}\right)
  +\left[\frac{1}{4}
    +\frac{2\mgluino\mtop \sin 2\theta_t - \mgluino^2 - \mtop^2}
    {4\mstop{1}^2}
  \right]B_0^{\rm fin}(\mstop{1}^2,\mtop,\mgluino)
  \Bigg\}
  \,.
\label{eq::mstpole}
\end{split}
\end{equation}
The corresponding relation for the mass $\mstop{2}$ is obtained by
interchanging the indices ``1'' and ``2'' and changing the sign of
$\sin 2\theta_t$.

Finally, for the mixing angle we have
\begin{eqnarray}
  \theta_t^\bare &=& \theta_t + \delta\theta_t\,,
  \nonumber\\
  \delta\theta_t &=& C_F\frac{\alpha_s}{\pi}
  \frac{\cos 2\theta_t}{\mstop{1}^2-\mstop{2}^2}
  \Bigg\{
  \frac{4\mgluino\mtop+(\mstop{2}^2-\mstop{1}^2)\sin 2\theta_t}
  {4\epsilon}
  \nonumber\\&&\mbox{}
  +\frac{\sin 2\theta_t}{4}\left[
    \mstop{2}^2\left(1+\Lstop{2}\right)
    -\mstop{1}^2\left(1+\Lstop{1}\right)
  \right]
  + \mgluino\mtop B_0^{\rm fin}(q_0^2,\mtop,\mgluino)
  \Bigg\}
  \,.
  \label{eq::deltatheta}
\end{eqnarray}
For $q^2\le(m_1-m_2)^2$, $B_0^{\rm fin}(q^2,m_1,m_2)$ is given by
\begin{eqnarray}
  B_0^{\rm fin}(q^2,m_1,m_2) &=&
  2-\ln\frac{m_1m_2}{\muR^2}
  + \frac{m_1^2-m_2^2}{q^2} \ln\frac{m_2}{m_1}
  + \frac{M_+ M_-}{q^2}
  \ln\frac{M_+ + M_-}{M_+ - M_-}
  \,,
\end{eqnarray}
with $M_\pm = \sqrt{(m_1\pm m_2)^2 -q^2}$.  The analytical expressions
for the other kinematical regions can be derived from this expression by
proper analytical continuation.  
Note that the counter terms in the $\drbar$ scheme are obtained by
discarding the finite parts at order $\alpha_s$.

The decoupling constant entering \eqn{eq::projector} is
defined in analogy to \eqn{eq::asdec} via the relation of the bare gluon
field of the full theory, $\tilde{G}_\mu^\bare$, (i.e. including the top quark
and the \susy{} particles) and the effective theory,
$G_\mu^\bare$:
\begin{eqnarray}
  G_\mu^\bare &=&  \sqrt{\zeta_3^\bare} \tilde{G}^\bare_\mu
  \,,
\end{eqnarray}
where
\begin{equation}
\begin{split}
\zeta_3^\bare &= 1 + \api\left[\frac{1}{\ep}
  \left(\frac{1}{6}C_A + \frac{1}{2}T\right)  
  + L(\ep)\right] +
\order{\alpha_s^2}\,,
\label{eq::zeta30}
\end{split}
\end{equation}
with $L(\ep)$ from \eqn{eq::l}.

$\alpha_s^\drbar$ is transformed from the $\drbar$ to the $\msbar$ scheme
through a finite shift~\cite{Martin:1993yx}; however, we found that this
shift is canceled by a finite shift in the decoupling constant
$\zeta_g^\bare$ and the operator renormalization $Z_{11}$, given in
\eqn{eq::opren}. Our final result is thus expressed in terms of
the $\msbar$ coupling $\alpha_s$ for standard
five-flavor \qcd{}, on-shell quark and squark masses, and the squark
mixing angle as defined in \eqn{eq::deltatheta} (the gluino mass is
unaffected by renormalization at the order considered here).

\section{Description of {\tt evalcsusy.f}}\label{sec::evalcsusy}

In this Appendix we give details on the usage of the {\tt Fortran} program
{\tt evalcsusy.f}~\cite{evalcsusy}. The distribution includes the files
\begin{verbatim}
 evalcsusy.f  common-slha.f  functions.f  readslha.f  slhablocks.f
\end{verbatim}
as well as
the {\tt Fortran} code for the two-loop
results of $C_1$ as defined in \eqn{eq::C1}; the latter is contained in
the directory {\tt obj/}.  For compilation one also needs the
{\abbrev CERN} libraries {\tt kernlib} and {\tt mathlib}.

In a first step, object files are generated with
\begin{verbatim}
> f77 -c functions.f readslha.f slhablocks.f obj/*.f
\end{verbatim}
The executable is then obtained by saying
\begin{verbatim}
> f77 -o evalcsusy evalcsusy.f functions.o readslha.o \
  slhablocks.o obj/*.o -L`cernlib -v pro kernlib,mathlib`
\end{verbatim}

{\tt evalcsusy} is invoked as
\begin{verbatim}
> ./evalcsusy <infile> <outfile>
\end{verbatim}
where {\tt <infile>} and {\tt <outfile>} are the in- and output file,
respectively. 
Both files obey the \susy{} Les Houches accord ({\abbrev
SLHA})~\cite{Skands:2003cj} which makes it straightforward to interface
{\tt evalcsusy.f} with a spectrum calculator.  The basic idea of the
{\abbrev SLHA} is to group the parameters into various blocks which have
a uniquely defined structure in order to ensure universality.  For
our process we need some parameters (the precise specification can be
seen in the example presented below) of the blocks {\tt SMINPUTS}, {\tt
MASS}, {\tt ALPHA}, {\tt HMIX}, {\tt STOPMIX} and {\tt MINPAR}.  In
addition, we introduce a new block {\tt CREIN} specific to {\tt
  evalcsusy.f}, where the ratio $\muR/M_h$ and
the parameter $q_0$ is defined (see \eqn{eq::q0} and \sct{sec::renorm}).  
The latter is actually composed out of
the three quantities $q_0^c$, $q_{01}$ and $q_{02}$ via the relation
\begin{eqnarray}
  q_0 = q_0^c + q_{01} \mstop{1} + q_{02} \mstop{2}
  \,.
\end{eqnarray}
If $q_{01}$ ($q_{02}$) is not defined, its value is set to zero.

{\tt evalcsusy.f} copies the contents of the input file to the output
file and appends an additional block {\tt HGGSUSY}. Its structure is as
follows:
\begin{description}
\item{{\tt Block HGGSUSY}}
  \begin{description}
  \item {\verb$101$} \hspace{.2em} : $c_1^{(0),\rm SM}$
  \item {\verb$102$} \hspace{.2em} : $c_1^{(1),\rm SM}$
  \item {\verb$103$} \hspace{.2em} : $c_1^{(2),\rm SM}$
  \item {\verb$201$} \hspace{.2em} : $c_1^{(0)}$
  \item {\verb$202$} \hspace{.2em} : $c_1^{(1)}$
  \item {\verb$1001$} : $g_t^h$
  \item {\verb$1011$} : $g_{t,11}^h$
  \item {\verb$1022$} : $g_{t,22}^h$
  \item {\verb$1012$} : $g_{t,12}^h$
  \item {\verb$1021$} : $g_{t,21}^h$
  \end{description}
\end{description}
It contains the results for the one- and two-loop coefficients
(c.f.\ \eqn{eq::C1}) both for the \sm{} ($c_1^{(0),\rm SM}$ and
$c_1^{(1),\rm SM}$) and the \mssm{} ($c_1^{(0)}$ and $c_1^{(1)}$).
Furthermore, the three-loop Standard Model term is given.
In addition, it provides 
numerical values for
the top and stop Yukawa couplings $g_t^h$ and $g_{t,ij}^h$ (see
Eqs.\,(\ref{eq::tophiggs})--(\ref{eq::stophiggs-1})).

Let us exemplify the use of {\tt evalcsusy.f} by considering the
{\abbrev SPS}\,1a point given in \eqn{eq::SPS1a}. Defining in
addition the \sm{} parameters of \eqn{eq::sminputs} in the block {\tt
SMINPUTS}, the output of {\tt SoftSusy}~\cite{Allanach:2001kg} looks as
follows (after adding the block \verb|CREIN|):
\begin{verbatim}
Block CREIN
 6  0.d0        # q0c
 61 .5d0        # q01
 62 .5d0        # q02
 7  1.d0        # renormalization scale muR/mh

Block SMINPUTS   # Standard Model inputs
     3    1.18000000e-01   # alpha_s(MZ)MSbar
     4    9.11876000e+01   # MZ(pole)
     6    1.78000000e+02   # Mtop(pole)

Block MINPAR  # SUSY breaking input parameters
     3    1.00000000e+01   # tanb
     4    1.00000000e+00   # sign(mu)
     1    1.00000000e+02   # m0
     2    2.50000000e+02   # m12
     5   -1.00000000e+02   # A0

Block MASS  # Mass spectrum
#PDG code      mass              particle
        24     8.02591534e+01   # MW
        25     1.12153306e+02   # h0
   1000006     3.97398225e+02   # ~t_1
   1000021     6.11147741e+02   # ~g
   2000006     5.86830420e+02   # ~t_2

Block alpha   # Effective Higgs mixing parameter
          -1.13348399e-01   # alpha

Block hmix Q= 4.661391312e+02  # Higgs mixing parameters
    1    3.65690378e+02  # mu

Block stopmix  # stop mixing matrix
 1  1     5.34006091e-01   # O_{11}
 1  2     8.45480617e-01   # O_{12}
\end{verbatim}
where only those parameters are displayed
which are needed in {\tt evalcsusy.f}. 
Using this file as input for {\tt evalcsusy.f}, its contents are copied
to the output file, and the block
\begin{verbatim}
Block HGGSUSY
     101    0.10000000E+01  #           cSM 1-loop
     102    0.27500000E+01  #           cSM 2-loop
     103    0.35160026E+01  #           cSM 3-loop
     201    0.10343231E+01  #         cSUSY 1-loop
     202    0.24377938E+01  #         cSUSY 2-loop
    1001    0.99853850E+00  #                  gth
    1011   -0.53035841E+00  #                gth11
    1022    0.42680033E+01  #                gth22
    1012    0.11983135E+01  #                gth12
    1021    0.11983135E+01  #                gth21
\end{verbatim}
is added.

\end{appendix}

\def\app#1#2#3{{Act.~Phys.~Pol.~}\jref{\bf B #1}{#2}{#3}}
\def\apa#1#2#3{{Act.~Phys.~Austr.~}\jref{\bf#1}{#2}{#3}}
\def\annphys#1#2#3{{Ann.~Phys.~}\jref{\bf #1}{#2}{#3}}
\def\cmp#1#2#3{{Comm.~Math.~Phys.~}\jref{\bf #1}{#2}{#3}}
\def\cpc#1#2#3{{Comp.~Phys.~Commun.~}\jref{\bf #1}{#2}{#3}}
\def\epjc#1#2#3{{Eur.\ Phys.\ J.\ }\jref{\bf C #1}{#2}{#3}}
\def\fortp#1#2#3{{Fortschr.~Phys.~}\jref{\bf#1}{#2}{#3}}
\def\ijmpc#1#2#3{{Int.~J.~Mod.~Phys.~}\jref{\bf C #1}{#2}{#3}}
\def\ijmpa#1#2#3{{Int.~J.~Mod.~Phys.~}\jref{\bf A #1}{#2}{#3}}
\def\jcp#1#2#3{{J.~Comp.~Phys.~}\jref{\bf #1}{#2}{#3}}
\def\jetp#1#2#3{{JETP~Lett.~}\jref{\bf #1}{#2}{#3}}
\def\jhep#1#2#3{{JHEP~}\jref{\bf #1}{#2}{#3}}
\def\mpl#1#2#3{{Mod.~Phys.~Lett.~}\jref{\bf A #1}{#2}{#3}}
\def\nima#1#2#3{{Nucl.~Inst.~Meth.~}\jref{\bf A #1}{#2}{#3}}
\def\npb#1#2#3{{Nucl.~Phys.~}\jref{\bf B #1}{#2}{#3}}
\def\nca#1#2#3{{Nuovo~Cim.~}\jref{\bf #1A}{#2}{#3}}
\def\plb#1#2#3{{Phys.~Lett.~}\jref{\bf B #1}{#2}{#3}}
\def\prc#1#2#3{{Phys.~Reports }\jref{\bf #1}{#2}{#3}}
\def\prd#1#2#3{{Phys.~Rev.~}\jref{\bf D #1}{#2}{#3}}
\def\pR#1#2#3{{Phys.~Rev.~}\jref{\bf #1}{#2}{#3}}
\def\prl#1#2#3{{Phys.~Rev.~Lett.~}\jref{\bf #1}{#2}{#3}}
\def\pr#1#2#3{{Phys.~Reports }\jref{\bf #1}{#2}{#3}}
\def\ptp#1#2#3{{Prog.~Theor.~Phys.~}\jref{\bf #1}{#2}{#3}}
\def\ppnp#1#2#3{{Prog.~Part.~Nucl.~Phys.~}\jref{\bf #1}{#2}{#3}}
\def\rmp#1#2#3{{Rev.~Mod.~Phys.~}\jref{\bf #1}{#2}{#3}}
\def\sovnp#1#2#3{{Sov.~J.~Nucl.~Phys.~}\jref{\bf #1}{#2}{#3}}
\def\sovus#1#2#3{{Sov.~Phys.~Usp.~}\jref{\bf #1}{#2}{#3}}
\def\tmf#1#2#3{{Teor.~Mat.~Fiz.~}\jref{\bf #1}{#2}{#3}}
\def\tmp#1#2#3{{Theor.~Math.~Phys.~}\jref{\bf #1}{#2}{#3}}
\def\yadfiz#1#2#3{{Yad.~Fiz.~}\jref{\bf #1}{#2}{#3}}
\def\zpc#1#2#3{{Z.~Phys.~}\jref{\bf C #1}{#2}{#3}}
\def\ibid#1#2#3{{ibid.~}\jref{\bf #1}{#2}{#3}}

\newcommand{\jref}[3]{{\bf #1}, #3 (#2)}
\newcommand{\bibentry}[4]{#1, {\it #2}, #3}
\newcommand{\arxiv}[1]{{\tt #1}}

\end{document}